\documentclass[11pt,british]{article}
\usepackage[T1]{fontenc}
\usepackage[latin9]{inputenc}
\usepackage[a4paper]{geometry}
\usepackage{babel}
\usepackage{float}
\usepackage{amsthm}
\usepackage{amsmath}
\usepackage{amssymb}
\usepackage{esint}
\usepackage[authoryear]{natbib}
\usepackage{lmodern}
\usepackage{graphics}
\usepackage{microtype}
\usepackage[unicode=true,pdfusetitle,
 bookmarks=true,bookmarksnumbered=false,bookmarksopen=false,
 breaklinks=false,pdfborder={0 0 0},backref=false,colorlinks=true]
 {hyperref}

\makeatletter

\theoremstyle{plain}
\newtheorem{conjecture}{Conjecture}

\newcommand{\includefigure}[1]{\centering\includegraphics{figures/#1}}

\def\rd{\mathrm{d}}
\def\supp{\operatorname{supp}}
\def\sign{\operatorname{sign}}
\def\sech{\operatorname{sech}}
\def\diag{\operatorname{diag}}
\def\bu{\boldsymbol{u}}
\def\bx{\boldsymbol{x}}
\def\bff{\boldsymbol{f}}
\def\bF{\boldsymbol{F}}
\def\bn{\boldsymbol{n}}
\def\bnabla{\boldsymbol{\nabla}}
\def\bcdot{\boldsymbol{\cdot}}
\def\bwedge{\boldsymbol{\wedge}}
\def\bzero{\boldsymbol{0}}

\makeatother

\begin{document}

\title{Asymptotic behaviour of solutions to the stationary\\
Navier-Stokes equations in two dimensional exterior domains\\
with zero velocity at infinity}

\author{\href{mailto:julien.guillod@unige.ch}{Julien Guillod} and \href{mailto:peter.wittwer@unige.ch}{Peter Wittwer}\\
{\small Department of Theoretical Physics,}\\
{\small University of Geneva, Switzerland}}

\maketitle

\begin{abstract}
We investigate analytically and numerically the existence of stationary
solutions converging to zero at infinity for the incompressible Navier-Stokes
equations in a two-dimensional exterior domain. More precisely, we
find the asymptotic behaviour for such solutions in the case where
the net force on the boundary of the domain is non-zero. In contrast
to the three dimensional case, where the asymptotic behaviour is given
by a scale invariant solution, the asymptote in the two-dimensional
case is not scale invariant and has a wake. We provide an asymptotic
expansion for the velocity field at infinity, which shows that, within
a wake of width $|\bx|^{2/3}$, the velocity decays like $|\bx|^{-1/3}$,
whereas outside the wake, it decays like $|\bx|^{-2/3}$. We check
numerically that this behaviour is accurate at least up to second
order and demonstrate how to use this information to significantly
improve the numerical simulations. Finally, in order to check the
compatibility of the present results with rigorous results for the
case of zero net force, we consider a family of boundary conditions
on the body which interpolate between the non-zero and the zero net
force case.
\end{abstract}
\textit{\small MSC class:}{\small{} 76D03, 76D05, 35Q30, 76D10, 76D25,
74F10, 76M10}{\small \par}

\section{Introduction}

In what follows, we discuss the question of the existence of solutions
for the incompressible Navier-Stokes equations in the exterior domain
$\Omega=\mathbb{R}^{2}\setminus B(\bzero,1)$ where $B(\bzero,1)$
is the closed disk of radius one centred at the origin,\begin{subequations}
\begin{equation}
\begin{aligned}\Delta\bu-\bnabla p & =\bu\bcdot\bnabla\bu\,, & \bnabla\bcdot\bu & =0\,,\\
\left.\bu\right|_{\partial B(\bzero,1)} & =\bu^{*}\,, & \lim_{|\bx|\to\infty}\bu & =\bzero\,,
\end{aligned}
\label{eq:ns-0}
\end{equation}
where $\bu^{*}$ is any smooth boundary condition with no net flux,
\begin{equation}
\int_{\partial B(\bzero,1)}\bu^{*}\bcdot\bn=0\,.\label{eq:no-flux}
\end{equation}
\label{eq:ns-noforce}\end{subequations}In these equations, $\bu$
is the velocity field, $p$ the pressure and $\bn$ the outward normal
unit vector to the body $B(\bzero,1)$. By using the scaling symmetry
of the Navier-Stokes equations, this also covers the case of a disk
of arbitrary size. Also, our results are very likely not changed if
one replaces the disk with an arbitrary bounded domain with a smooth
enough boundary. We define the net force $\bF$ and the torque $M$
acting on the body by
\begin{align}
\bF & =\int_{\partial B(\bzero,1)}\mathbf{T}\bn\,, & M & =\int_{\partial B(\bzero,1)}\bx\bwedge\mathbf{T}\bn\,,\label{eq:FM-body}
\end{align}
where $\mathbf{T}$ is the stress tensor including the convective
part, $\mathbf{T}=\bu\otimes\bu+p-\bnabla\bu-\left(\bnabla\bu\right)^{T}$.
The net force and the torque are the two conserved quantities of the
Navier-Stokes equations, in the sense that \eqref{eq:FM-body} is
invariant if one replaces $\partial B(\bzero,1)$ by any homotopic
smooth curve. In particular the curve can be pushed to infinity, so
that these two quantities are encoded in the behaviour of the solution
at infinity, which plays an important role below.

Problem \eqref{eq:ns-noforce} is closely related to the one of the
incompressible Navier-Stokes equations in $\mathbb{R}^{2}$,
\begin{align}
\Delta\bu-\bnabla p & =\bu\bcdot\bnabla\bu-\bff\,, & \bnabla\bcdot\bu & =0\,, & \lim_{|\bx|\to\infty}\bu & =\bzero\,,\label{eq:ns-force}
\end{align}
where the force term $\bff$ is a smooth function of compact support.
See for example \citet{Hillairet.Wittwer-Asymptoticdescriptionof2011}
where the connection between two such problems is made precise. For
problem \eqref{eq:ns-force}, the net force $\bF$ and the torque
$M$ are defined by
\begin{align}
\bF & =\int_{\mathbb{R}^{2}}\bff\,, & M & =\int_{\mathbb{R}^{2}}\bx\bwedge\bff\,.\label{eq:FM-domain}
\end{align}

Even for small data $\bu^{*}$ or $\bff$, it is an open question
if problems \eqref{eq:ns-noforce} and \eqref{eq:ns-force} admit
a solution for general data \citep[see][for a complete review of the question]{Galdi-IntroductiontoMathematical2011,Galdi-StationaryNavier-Stokesproblem2004},
and the resolution of this question is one of the most challenging
open mathematical problems of two-dimensional stationary fluid mechanics
\citep{Yudovich-ElevenGreatProblems2003}. The main difficulty is
to determine the behaviour of solutions at infinity: the existence
of so-called $D$-solutions for the Navier-Stokes equations is a well-known
result, but the function spaces used in these proofs are not sufficiently
restrictive to prove that the velocity goes to zero at infinity.

The difficulty of proving that the boundary condition at infinity
is satisfied is specific to the two-dimensional stationary Navier-Stokes
equations with zero velocity at infinity \citep[see][notes on \S XII.5]{Galdi-IntroductiontoMathematical2011}.
More precisely, this difficulty is related to the fact that the linearisation
around zero of the Navier-Stokes equation is the Stokes equation,
and it is known that solutions of the Stokes equation in two-dimensions
do in general not decay at infinity. This fact is known as the Stokes
paradox. If solutions to the Navier-Stokes equations do exist, their
asymptotic behaviour can therefore not in general be given by a solution
of the Stokes equation, and this is why the question of the existence
of such solutions relies heavily on guessing the correct asymptotic
behaviour.

We note that the problem which one obtains when the boundary condition
at infinity is replaced by a non-zero constant vector field is completely
different from the one considered here \citep{Finn-stationarysolutionsNavier1967}.
The linearisation around the constant vector field at infinity leads
to the Oseen equation which, in contrast to the Stokes equation, possesses
a fundamental solution which decays to zero at infinity. Therefore,
a fixed point argument can be used in that case \citep{Galdi.Sohr-asymptoticstructureof1995,Sazonov-AsymptoticBehaviorof1999},
and the solution of the Navier-Stokes equations is asymptotic at infinity
to the Oseen fundamental solution.

In three dimensions, the situation is somewhat similar, since there
is also a fundamental distinction between the case where the velocity
at infinity is zero, or a non-zero constant vector field. However,
in contrast to the two-dimensional case, the functions spaces used
for the construction of $D$-solutions ensure in three dimensions
that the velocity converges to the prescribed value at infinity \citep[see][\S X.4]{Galdi-IntroductiontoMathematical2011}.
For a non-zero constant vector field at infinity, the relevant linear
problem is again the Oseen equation and the asymptotic behaviour of
the velocity is given by the Oseen fundamental solution \citep{Finn-Estimatesatinfinity1959,Babenko-stationarysolutionsof1973,Galdi-AsymptoticStructureDsol-1992,Farwig.Sohr-WeightedestimatesOseen1998}.
In the case of zero velocity at infinity, \citet{Galdi-asymptoticPropertiesLerays1993}
proves that the solution decays at infinity as $|\bx|^{-1}$, \emph{i.e.}
like the fundamental solution of the three-dimensional Stokes equation.
The solution, nevertheless, does not admit the Stokes fundamental
solution as its asymptote \citep{Deuring.Galdi-AsymptoticBehaviorof2000}.
The reason for this is that a solution decaying like $|\bx|^{-1}$
makes the linear term of the Navier-Stokes equation, $\Delta\bu$,
and the non-linear one, $\bu\bcdot\bnabla\bu$, having both the same
decay like $|\bx|^{-3}$. Recently \citet{Nazarov.Pileckas-AsymptoticofSolutions1999,Nazarov-steady2000}
showed that the asymptotic behaviour is given by a self-similar solution
decaying like $|\bx|^{-1}$, and \citet{Korolev.Sverak-largedistanceasymptotics2011}
then proved that the asymptotic behaviour coincides with solutions
of the Navier-Stokes equations in $\mathbb{R}^{3}\setminus\left\{ \bzero\right\} $
found by Landau \citep{Landau-newexactsolution1944}. The family of
Landau solutions depends on one real parameter which is related to
the net force, and therefore the asymptote encodes the information
concerning the net force at infinity. In two dimensions, the analogous
self-similar solutions are given by Hamel solutions \citep{Hamel-SpiralfoermigeBewegungen1917},
but in contrast to the three-dimensional case, the Hamel solutions
depend on a discrete parameter, and therefore cannot encode the net
force at infinity. Indeed, \citet[\S 5]{Sverak-LandausSolutionsNavier2011}
proves that the asymptote in two dimensions cannot be a self-similar
solution and in particular that one cannot obtain a solution to \eqref{eq:ns-force}
with perturbation techniques in a space a functions that decay like
$O(|\bx|^{-1})$ at infinity. In any case, solutions decaying like
$O(|\bx|^{-1})$ at infinity decay too fast to encode a non-zero net
force at infinity.

For all these reasons, the two-dimensional case with zero velocity
at infinity is particularly difficult and remains the only stationary
case where the existence of solutions satisfying the boundary condition
at infinity is not known for general data. A few results are available
under symmetry assumptions on the data, by \citet[\S 3.3]{Galdi-StationaryNavier-Stokesproblem2004}
and by \citet[\S 4.4]{Russo-Steady-StateNavier-StokesEquations2008}
for the body case \eqref{eq:ns-noforce}, and by \citet{Yamazaki-stationaryNavier-Stokesequation2009}
for the case of a force \eqref{eq:ns-force}. \citet{Pileckas-existencevanishing2012}
also consider symmetric boundary conditions, but allow a non-zero
net flux through the boundary. In all cases, the symmetry assumptions
imply either a zero net force on the boundary or a zero mean of the
force $\bff$. Recently, \citet{Hillairet-mu2013} proved the existence
of solutions decaying like $|\bx|^{-1}$ for a ball of boundary conditions
that are not centred at zero; more precisely the ball is centred on
the non-slip boundary condition corresponding to a rotating body.

Note that the analogy between the problem \eqref{eq:ns-noforce} and
\eqref{eq:ns-force} is at a formal level only, since in order to
make the analogy precise, existence and uniqueness of solutions needs
to be known in both cases. Formally, to pass from a solution for the
source force case \eqref{eq:ns-force} to a solution of the body case
\eqref{eq:ns-noforce}, one simply evaluates the solution $\bu$ on
$\partial\Omega$ which provides the corresponding $\bu^{*}$ (we
assume here that $\supp\bff\subset B(\bzero,1)$, which can always
been achieved by the scale invariance of the Navier-Stokes equations).
Conversely, one can cutoff the stream function of the body problem
to obtain a solution in the whole space with a source force. In this
step, the relation \eqref{eq:no-flux} is crucial to ensure that the
stream function in the exterior domain exists globally. As mentioned
above, the equivalence has been proven for an analogue problem with
constant velocity at infinity \citep{Hillairet.Wittwer-Asymptoticdescriptionof2011}.

The present work, is a first step towards solving the open problems
\eqref{eq:ns-noforce} and \eqref{eq:ns-force} for the case of a
non-zero net force $\bF$. Guided by related problems \citep{Dyke-Perturbationmethodsin1975}
and in particular by the semi-infinite plate problem \citep{Goldstein-LecturesFluidMechanics1957,Ockendon.Ockendon-ViscousFlow1995,Bichsel.Wittwer-Stationaryflowpast2007},
we look for an asymptotic expansion describing the asymptotic behaviour
at large distances from the origin. As explained below, the Navier-Stokes
equations together with the condition of a non-zero net force fixes
the decay of the velocity within the wake: if the decay is too slow,
the force is infinite, and if the decay is too fast, the force is
zero. Moreover, we show below that the requirement of the torque to
be finite implies that the first two orders of the asymptotic expansion
are symmetric with respect to an axis aligned in the direction the
net force $\bF$. We have the following conjecture:
\begin{conjecture}
\label{conj:main}For a large class of boundary conditions $\mathbf{u}^{*}$
(resp. source terms $\bff$) with a non-zero net force $\bF$, there
exists a solution to \eqref{eq:FM-body} (resp. to \eqref{eq:ns-force})
which satisfies 
\begin{align*}
u & =u_{0}+u_{1}+O(r^{-1})\,, & p & =p_{0}+p_{1}+O(r^{-2})\,,\\
v & =v_{0}+v_{1}+O(r^{-4/3})\,, & \omega & =\omega_{0}+\omega_{1}+O(r^{-5/3})\,,
\end{align*}
where $\bu=\left(u,v\right)$, $\omega=\bnabla\bwedge\bu$, and $r=|\bx|$.
For $i\in\left\{ 0,1\right\} $, the functions defining the asymptotic
behaviour satisfy $u_{i}=O(r^{-(1+i)/3})$, $v_{i}=O(r^{-(2+i)/3})$,
$p_{i}=O(r^{-(4+i)/3})$ and $\omega_{i}=O(r^{-(3+i)/3})$, and moreover
depend only on the net force $\bF$. In a coordinate system where
$\bF=\left(F,0\right)$ with $F>0$, the asymptotes are given explicitly,
for $i=0$, by \eqref{eq:order0} with $b_{0}=0$, and, for $i=1$,
by \eqref{eq:order1} with $b_{1}=0$. The parameter $a$ in the asymptotes
is linked to the net force $F$ through \eqref{eq:link-a-F}.
\end{conjecture}
In order to check that the asymptotic expansion at first \eqref{eq:order0}
and second \eqref{eq:order1} order describes the behaviour at infinity
correctly, we solve the problem numerically on truncated domains of
increasing size and look at the decay of the horizontal velocity $u$
up-stream and down-stream along the $x$-axis, as well as at the profile
of $u$ and $\omega$ in the $y$-direction at fixed values of $x$.
We then show that the knowledge of the asymptotic behaviour can significantly
improve the numerical simulations, by using the asymptotic terms to
define an artificial boundary condition \citep{Boenisch.etal-Adaptiveboundaryconditions2005,Boeckle.Wittwer-Artificialboundaryconditions2013}.
Finally, in view of the solutions found by \citet{Hillairet-mu2013}
we study the transition from the case of non-zero net force to zero
force, by varying the boundary condition on the body appropriately.

\section{Asymptotic expansion}

In this section, we limit the discussion to problem \eqref{eq:ns-force}
of a force of compact support, since the analysis is identical for
problem \eqref{eq:ns-noforce}. We only treat the case where the net
force $\bF$ defined by \eqref{eq:FM-domain} is different from zero.
Without loss of generality we choose the orientation of the coordinates
$\bx=\left(x,y\right)$ such that $\bF=\left(F,0\right)$, with $F>0$.
Moreover, modulo a translation in the $y$-direction, we can always
choose the coordinates such that the torque $M$ defined in \eqref{eq:FM-domain}
is zero, $M=0$.

In order to simplify the calculations, we choose to work with the
vorticity equation,
\begin{align}
\Delta\omega-\bu\bcdot\bnabla\omega & =-\bnabla\bwedge\bff\,, & \bnabla\bcdot\bu & =0\,,\label{eq:eq-omega}
\end{align}
where $\omega=\bnabla\bwedge\bu$ is the vorticity, since that way
the pressure is eliminated. The incompressible vorticity equation
\eqref{eq:eq-omega} reduces to a partial differential equation of
order four for the stream function $\psi$, which is defined by $\bu=\bnabla\wedge\psi$.
Once the stream function as been determined, we can construct the
pressure $p$ by integrating, for example, the second component of
the Navier-Stokes equation with respect to $y$. In what follows,
we use the stream function as the fundamental quantity in order to
find the asymptotic behaviour: we make an Ansatz for the stream function
which includes a wake, find the correct parameters of the wake based
on physical arguments and compute the first two leading terms of the
expansion.

Since the net force is oriented along the $x$-axis, it is natural
to discuss the symmetries of the Navier-Stokes equations \eqref{eq:ns-force}
with respect to the horizontal axis. It is well known that the equation
for the stream function is invariant under the symmetry given by $\psi(x,y)=-\psi(x,-y)$,
provided that the force satisfies $\bff(x,y)=\diag(1,-1)\bff(x,-y)$,
which in \eqref{eq:ns-force} corresponds to the symmetry, $\bu(x,y)=\diag(1,-1)\bu(x,-y)$
and $p(x,y)=-p(x,-y)$. Such solutions will be called symmetric in
what follows.

\subsection{Wake parameters}

Since $F>0$, it is natural \citep[see][]{Goldstein-LecturesFluidMechanics1957,Dyke-Perturbationmethodsin1975,Bichsel.Wittwer-Stationaryflowpast2007,Boenisch.etal-Secondorderadaptive2008}
to look for a wake region in the half-plane characterized by $x>0$
with a wake variable defined by $z=y/x^{p}$, where $0<p<1$. We first
consider only the wake region $x>0$, where in all known cases the
velocity field has the slowest decay. Below we then construct the
stream function in the whole plane by multiplying the functions describing
the wake by a Heaviside function $H(x)$, and by adding a harmonic
function to restore all the boundary conditions \citep[see][]{Goldstein-LecturesFluidMechanics1957,Bichsel.Wittwer-Stationaryflowpast2007}.
We start with the following Ansatz for the dominant term of the stream
function in the wake, \emph{i.e.} for large $x$ at fixed $z$,
\[
\psi(x,y)\approx x^{q}\varphi_{0}(z)+o(x^{q})\,,
\]
where $0<p<1$ and $q\in\mathbb{R}$. Under the usual assumptions
concerning the differentiability of the asymptotic expansion \citep{Goldstein-LecturesFluidMechanics1957,Dyke-Perturbationmethodsin1975},
the velocity field $\bu=\left(u,v\right)$ and the vorticity $\omega=\bnabla\bwedge\bu$
admit the expansions
\begin{align*}
u(x,y) & =x^{q-p}\varphi_{0}^{\prime}(z)+o(x^{q-p})\,,\\
v(x,y) & =x^{q-1}\left(pz\varphi_{0}^{\prime}(z)-q\varphi_{0}(z)\right)+o(x^{q-1})\,,\\
\omega(x,y) & =-x^{q-2p}\varphi_{0}^{\prime\prime}(z)+o(x^{q-2p})\,.
\end{align*}
By plugging this Ansatz into the equation for the vorticity we obtain
\begin{equation}
\Delta\omega-\bu\bcdot\bnabla\omega=-x^{q-4p}\varphi_{0}^{(4)}(z)-x^{2q-3p-1}\left(q\varphi_{0}(z)\varphi_{0}^{(3)}(z)+\left(2p-q\right)\varphi_{0}^{\prime}(z)\varphi_{0}^{\prime\prime}(z)\right)+o(x^{\max(q-4p,2q-3p-1)})\,.\label{eq:eq-omega-ans}
\end{equation}
In order to obtain a differential equation for the wake involving
the linear and non-linear part of the Navier-Stokes equation, we have
to choose
\begin{equation}
q-4p=2q-3p-1\qquad\Longleftrightarrow\qquad p+q=1\,.\label{eq:pq-1}
\end{equation}
With this condition the vorticity equation becomes
\begin{equation}
\Delta\omega-\bu\bcdot\bnabla\omega=-x^{1-5p}\left(\varphi_{0}^{(4)}(z)+\left(1-p\right)\varphi_{0}(z)\varphi_{0}^{(3)}(z)+\left(3p-1\right)\varphi_{0}^{\prime}(z)\varphi_{0}^{\prime\prime}(z)\right)+o(x^{1-5p})\,.\label{eq:eq-omega-o1}
\end{equation}
Since we are interested in a solution that provides a net force, we
need to satisfy the equation 
\begin{equation}
\bF=\left(F,0\right)=\lim_{R\to\infty}\lim_{S\to\infty}\int_{\partial\left(\left[-R,R\right]\times\left[-S,S\right]\right)}\mathbf{T}\bn\,,\label{eq:force-square}
\end{equation}
where $\mathbf{T}$ is the stress tensor including the convective
part. By assuming that the pressure and the velocity outside the wake
do not influence the net force, a fact that will be verified later
on, we find
\begin{equation}
\bF=\left(F,0\right)=\lim_{x\to\infty}\int_{-\infty}^{+\infty}\bu u\,\rd y=\lim_{x\to\infty}\left[\int_{-\infty}^{+\infty}\left(x^{2q-p}\varphi_{0}^{\prime}(z)^{2},0\right)\rd z+o(x^{2q-p})\right]\,.\label{eq:force-pq}
\end{equation}
In order to obtain a finite non-zero net force we therefore have to
choose $p$ and $q$ such that $2q-p=0$, which together with the
relation \eqref{eq:pq-1} fixes the exponents,
\begin{align}
p & =\frac{2}{3}\,, & q & =\frac{1}{3}\,,\label{eq:exponents}
\end{align}
and from \eqref{eq:force-pq} we therefore get that the net force
is related to the function $\varphi_{0}$ by
\begin{equation}
F=\int_{-\infty}^{+\infty}\varphi_{0}^{\prime}(z)^{2}\rd z\,.\label{eq:net-forcex}
\end{equation}

The previous analysis demonstrates that the stream function $\psi=x^{1/3}\varphi_{0}(z)$
with $\varphi_{0}$ the solution of the differential equation appearing
in the right hand side of \eqref{eq:eq-omega-o1}, satisfies the vorticity
equation \eqref{eq:eq-omega-o1} in the wake with a remainder of order
$O(x^{-8/3})$ for large $x$ at fixed $z$. The aim of the asymptotic
expansion which we construct now is to improve the decay of the remainder
at each step of the development. In view of \eqref{eq:eq-omega-ans}
and by requiring that the terms generated by the previous orders cancel
in the wake, the natural Ansatz for the stream function in the wake
is 
\begin{equation}
\psi=x^{1/3}\varphi_{0}(z)+\varphi_{1}(z)+x^{-1/3}\varphi_{2}(z)+\cdots\,,\label{eq:expansion}
\end{equation}
where $z=y/x^{2/3}$ is the wake variable. By plugging this Ansatz
into the vorticity equation \eqref{eq:eq-omega}, the terms of order
$x^{-(7+i)/3}$ define an ordinary differential equation for $\varphi_{i}$.
Consequently, the asymptotic expansion is in powers of $x^{-1/3}$,
eventually with logarithmic corrections entering the expansion at
some point \citep[see for example][\S 2]{Boenisch.etal-Secondorderadaptive2008}.
In the following we solve these differential equations for the first
two orders, and extend the asymptotic expansion from the wake region
to the whole plane by using techniques from harmonic analysis.

\subsection{First order term}

With $p$ and $q$ fixed by \eqref{eq:exponents}, the function $\varphi_{0}$
has to satisfy the differential equation obtained from \eqref{eq:eq-omega-o1},
\[
\varphi_{0}^{(4)}+\frac{1}{3}\varphi_{0}\,\varphi_{0}^{(3)}+\varphi_{0}^{\prime}\,\varphi_{0}^{\prime\prime}=0\,.
\]
After three explicit integrations, the equation becomes
\[
\varphi_{0}^{\prime}+\frac{1}{6}\varphi_{0}^{2}=C_{2}z^{2}+C_{1}z+C_{0}\,,
\]
where $C_{0},C_{1},C_{2}\in\mathbb{R}$ are some constants. The derivative
$\varphi_{0}^{\prime}$ has to be zero at infinity so that the horizontal
velocity $u$ satisfies the boundary condition \eqref{eq:ns-0} at
infinity, and the requirement that the velocity field is bounded implies
that $\varphi_{0}$ is bounded. This is only possible if $C_{1}=C_{2}=0$,
and consequently, the resulting bounded solutions are given explicitly
by
\begin{equation}
\varphi_{0}(z)=6a\tanh(az-b_{0})\,,\label{eq:def-phi0}
\end{equation}
where $a>0$ is a constant related to $C_{0}$ by $C_{0}=6a^{2}$,
and $b_{0}\in\mathbb{R}$. The parameter $b_{0}$ introduces symmetry
breaking, since for $b_{0}=0$, the stream function defined above
corresponds to a symmetric solution, whereas for $b_{0}\neq0$, the
solution is not symmetric. At the end of this subsection, we show
that $b_{0}=0$, since otherwise the torque would be infinite. We
note that $\varphi_{0}^{\prime}$ decays exponential fast at infinity,
and that $\varphi_{0}$ is bounded but does not converge to zero at
infinity. Therefore, as in the semi-infinite plate case \citep{Goldstein-LecturesFluidMechanics1957,Bichsel.Wittwer-Stationaryflowpast2007},
the horizontal component $u$ satisfies the boundary condition at
infinity, but $v$ does not go to zero at infinity for fixed $x$,
\begin{align*}
\lim_{y\to\pm\infty}u(x,y) & =\frac{1}{x^{1/3}}\lim_{z\to\pm\infty}\varphi_{0}^{\prime}(z)=0\,,\\
\lim_{y\to\pm\infty}v(x,y) & =\frac{1}{3x^{2/3}}\lim_{z\to\pm\infty}\left(2z\varphi_{0}^{\prime}(z)-\varphi_{0}(z)\right)=\frac{\mp1}{3x^{2/3}}\,.
\end{align*}
Following \citet{Goldstein-LecturesFluidMechanics1957} and others
\citep{Dyke-Perturbationmethodsin1975,Bichsel.Wittwer-Stationaryflowpast2007},
in order to restore the boundary condition for $v$, we introduce
a branch-cut along the positive real axis, subtract from $\varphi_{0}(z)$
its asymptote at large $z$, \emph{i.e.} $6a\sign z$, and compensate
the resulting discontinuity on the branch-cut by adding to the stream
function an appropriate harmonic function. Explicitly we take
\[
\psi_{0}=H(x)x^{1/3}\left(\varphi_{0}(z)-6a\sign z\right)-4\sqrt{3}ar^{1/3}\sin(\theta/3)\,,
\]
where $H$ is the Heaviside function, $r=\sqrt{x^{2}+y^{2}}$ and
$\theta=\arg\left(-x-\mathrm{i}y\right)$ is the angle measured from
the negative real axis. That way, and since $\varphi_{0}(z)-6a\sign z$
decays exponential fast at infinity, the term multiplied by $H(x)$
and the harmonic term are smooth in $\mathbb{R}^{2}\setminus\left[0,\infty\right)$\textcolor{magenta}{{}
}and the resulting velocity field decays at infinity. Moreover, the
function $\psi_{0}$ is by construction continuous across the branch-cut
and therefore continuous in the whole plane apart from the origin.
By adding higher order terms to $\psi_{0}$, one could obtain a stream
function which is arbitrarily smooth away from the origin. For simplicity
these terms are not written here, since they do not change the following
argument in any way. The first order of the velocity field and of
the vorticity are\begin{subequations}
\begin{align}
u_{0} & =\frac{H(x)}{x^{1/3}}\varphi_{0}^{\prime}(z)-\frac{4a}{\sqrt{3}}\frac{x\cos(\theta/3)+y\sin(\theta/3)}{r^{5/3}}\,,\label{eq:order0-u}\\
v_{0} & =\frac{H(x)}{3x^{2/3}}\left(2z\varphi_{0}^{\prime}(z)-\varphi_{0}(z)+6a\sign z\right)+\frac{4a}{\sqrt{3}}\frac{x\sin(\theta/3)-y\cos(\theta/3)}{r^{5/3}}\,,\label{eq:order0-v}\\
\omega_{0} & =-\frac{H(x)}{x}\varphi_{0}^{\prime\prime}(z)\,.\label{eq:order0-w}
\end{align}
Finally, by integrating the second component of the Navier-Stokes
equation with respect to $y$, and requiring that the pressure is
zero at infinity, we reconstruct the pressure term at leading order,
\begin{equation}
p_{0}=\frac{H(x)}{x^{4/3}}\frac{4a^{2}}{3}\left(\rho_{0}(z)-4b_{0}\sign z\right)+\frac{8a^{2}}{3r^{4/3}}+\frac{32a^{2}b_{0}}{3\sqrt{3}}\frac{\sin(4\theta/3)}{r^{4/3}}\,,\label{eq:order0-p}
\end{equation}
\label{eq:order0}\end{subequations}where
\[
\rho_{0}(z)=4az\tanh(az-b_{0})-4\log(2\cosh(az-b_{0}))+2az\tanh(az-b_{0})\text{sech}^{2}(az-b_{0})+7\text{sech}^{2}(az-b_{0})\,.
\]

As mentioned before, it is possible to smooth $\psi_{0}$ across the
branch-cut by adding a term of order $O(r^{-1/3})$ in a way that
all the previous leading terms are unaffected, and therefore this
smoothed version satisfies the Navier-Stokes system \eqref{eq:ns-force}
in the classical sense with a remainder $\bff_{0}=(O(r^{-2}),O(r^{-7/3}))$.

We now calculate the force $\bF$ and the torque $M$ at leading order.
To this end, we first compute the stress tensor including the convective
term at first order, \emph{i.e.} with a remainder of order $O(r^{-4/3})$,\begingroup\renewcommand*{\arraystretch}{1.4}
\[
\mathbf{T}_{0}=H(x)\begin{pmatrix}\frac{1}{x^{2/3}}\varphi_{0}^{\prime}(z)^{2} & \frac{2}{3x}z\varphi_{0}^{\prime}(z)^{2}\\
\frac{2}{3x}z\varphi_{0}^{\prime}(z)^{2} & 0
\end{pmatrix}\,.
\]
\endgroup Since $\varphi_{0}^{\prime}$ decays to zero at infinity,
the contribution of the integrals on the upper and lower lines of
the square $\left[-R,R\right]\times\left[-S,S\right]$ to the force
in \eqref{eq:force-square} are zero in the limit $S\to\infty$, so
that the force is given by
\[
\bF=\lim_{R\to\infty}\int_{-\infty}^{+\infty}\left[\left.\mathbf{T}\boldsymbol{e}_{1}\right|_{x=R}-\left.\mathbf{T}\boldsymbol{e}_{1}\right|_{x=-R}\right]\rd y\,,
\]
where $\boldsymbol{e}_{1}=\left(1,0\right)$ is the unit vector in
the $x$-direction. In consequence of that only the non-linear term
in the wake contributes to the force,
\begin{equation}
\bF=\lim_{x\to\infty}\left[\int_{-\infty}^{+\infty}\left(\varphi_{0}^{\prime}(z)^{2},0\right)\rd z+O(x^{-1/3})\right]=\left(48a^{3},0\right)\,.\label{eq:F-a}
\end{equation}
The torque at infinity is given by
\[
M=\lim_{R\to\infty}\lim_{S\to\infty}\int_{\partial\left(\left[-R,R\right]\times\left[-S,S\right]\right)}\mathbf{x}\bwedge\mathbf{T}\bn\,,
\]
and again the contributions from the horizontal boundaries vanish
in the limit $S\to\infty$, and we get
\begin{equation}
M=\lim_{R\to\infty}\int_{-\infty}^{+\infty}\left[\left.\bx\bwedge\mathbf{T}\boldsymbol{e}_{1}\right|_{x=R}-\left.\bx\bwedge\mathbf{T}\boldsymbol{e}_{1}\right|_{x=-R}\right]\rd y\,.\label{eq:M-limR}
\end{equation}
By explicit integration, we find that
\[
M=\lim_{x\to\infty}\left[\frac{-x^{2/3}}{3}\int_{-\infty}^{+\infty}z\varphi_{0}^{\prime}(z)^{2}\rd z+O(x^{1/3})\right]=\lim_{x\to\infty}\left[16a^{2}b_{0}x^{2/3}+O(x^{1/3})\right]\,,
\]
and therefore, in order for the torque to be finite, we have to set
$b_{0}=0$. Consequently the first order asymptote is symmetric, and
the torque at first order is zero. The contribution to the torque
hidden in the term $O(x^{1/3})$ is discussed in the next subsection.

\subsection{Second order term}

As explained above, the natural Ansatz for the second order term of
the stream function in the wake is $\psi_{1}=\varphi_{1}(z)$. As
for the first order term, we need to restore the boundary conditions
by adding a harmonic term, so we directly make the following Ansatz
for the second order term of the stream function in $\mathbb{R}^{2}\setminus\left\{ \boldsymbol{0}\right\} $,
\[
\psi_{1}=H(x)\left(\varphi_{1}(z)-\gamma\sign z\right)-\frac{\gamma}{\pi}\theta\,,
\]
where $\gamma$ is the limit of $\varphi_{1}$ at infinity which will
be determined later. As above, the harmonic function is introduced
to compensate the discontinuity coming from the term $\sign z$ which
is subtracted from $\varphi_{1}(z)$ in order to satisfy the boundary
condition of $v$ at infinity. We note that the harmonic function
in $\psi_{1}$ corresponds to a radial source at the origin and ensures
that the solution has no flux at infinity. In the case where \eqref{eq:no-flux}
is violated, \emph{i.e.} when there is a non-zero net flux, the asymptotic
behaviour of the solution is probably the same as the one constructed
here, except that the parameter $\gamma$ will be changed to allow
for a non-zero net flux.

By plugging $\psi_{0}+\psi_{1}$ into the vorticity equation, and
using the definition of $\varphi_{0}$, the leading term at large
$x$ and constant $z$ is given by the following linear equation for
$\varphi_{1}$,
\[
\varphi_{1}^{(4)}+\frac{1}{3}\varphi_{0}\,\varphi_{1}^{(3)}+\frac{4}{3}\varphi_{0}^{\prime}\,\varphi_{1}^{\prime\prime}+\varphi_{0}^{\prime\prime}\,\varphi_{1}^{\prime}=\frac{2a\varphi_{0}^{\prime\prime}}{\sqrt{3}}\,.
\]
The function $\varphi_{1}=\varphi_{0}^{\prime}$ is a solution of
the homogeneous equation, and consequently we can reduce the order
of the differential equation by one, and the resulting equation can
be solved explicitly. The general solution of the differential equation
for $\varphi_{1}$ which doesn't diverge at infinity is
\[
\varphi_{1}(z)=\frac{1}{2a\sqrt{3}}\left(z\varphi_{0}^{\prime}(z)+\varphi_{0}(z)\right)+b_{1}\varphi_{0}^{\prime}(z)=\sqrt{3}\left(\tanh(az)+az\sech^{2}(az)\right)+b_{1}\varphi_{0}^{\prime}(z)\,,
\]
with $b_{1}\in\mathbb{R}$. We find $\gamma=\sqrt{3}$ so that $\varphi_{1}(z)-\gamma\sign z$
converges to zero at infinity. For the second order terms of the velocity
field and of the vorticity, we get\begin{subequations}
\begin{align}
u_{1} & =\frac{H(x)}{x^{2/3}}\varphi_{1}^{\prime}(z)-\frac{\sqrt{3}}{\pi}\frac{x}{r^{2}}\,,\\
v_{1} & =\frac{2H(x)}{3x}z\varphi_{1}^{\prime}(z)-\frac{\sqrt{3}}{\pi}\frac{y}{r^{2}}\,,\\
\omega_{1} & =-\frac{H(x)}{x^{4/3}}\varphi_{1}^{\prime\prime}(z)\,.
\end{align}
The term multiplied by $b_{1}$ breaks the symmetry since all other
terms in the stream function are odd in the variable $y$. In fact,
as we show at the end of this subsection, the parameter $b_{1}$ also
has to be taken equal to zero, since otherwise the torque is infinite.
By integrating the second component of the Navier-Stokes equation
one recovers the pressure at second order,
\begin{equation}
p_{1}=\frac{H(x)}{x^{5/3}}\frac{2a}{3\sqrt{3}}\rho_{1}(z)+\frac{4a}{\pi}\frac{\cos(\theta/3)}{r^{5/3}}\,,\label{eq:order1-p}
\end{equation}
\label{eq:order1}\end{subequations}where
\[
\rho_{1}(z)=\text{sech}^{4}(az)\left(6a^{2}z^{2}-4az\sinh(2az)+7\cosh(2az)+7\right)\,.
\]
In these expressions, we have already suppressed the contributions
due to $b_{1}$ since we prove below that $b_{1}=0$. Again, it is
possible to smooth the sum of the first and of the second terms without
modifying these asymptotes, so that the Navier-Stokes system \eqref{eq:ns-force}
is satisfied with a remainder $\bff_{1}=(O(r^{-7/3}),O(r^{-8/3}))$.

As anticipated above, the contribution of the second order terms to
the net force is zero. For the torque, we find that \eqref{eq:M-limR}
is given at second order by
\[
M=\lim_{x\to\infty}\left[32a^{3}b_{1}x^{1/3}+O(x^{0})\right]\,,
\]
and therefore, in order to have a finite torque, we have to set $b_{1}=0$.
This implies that the first and the second order asymptotes are symmetric,
and that the torque generated by the first two asymptotic terms is
zero.

\subsection{Third order term}

It is possible to construct a third order term of the asymptotic expansion
\eqref{eq:expansion} with the Ansatz $\psi_{2}=x^{-1/3}\varphi_{2}(z)$
within the wake. The solution is given in term of Legendre functions.
The homogeneous differential equation for $\varphi_{2}$ admits the
solution $\varphi_{2}=\varphi_{0}^{\prime}$, which corresponds to
the generator of translations along the $y$-axis. As expected from
the previous calculations, the third order contribution to the torque
is finite. More precisely, the torque $M$ is given by the parameter
$b_{2}$ multiplying $\varphi_{0}^{\prime}$ in $\varphi_{2}$, in
complete agreement with the fact that we can set $M=0$ by a translation
in the $y$-direction. This third order computation justifies the
statement of conjecture~\ref{conj:main} concerning the decay of
the remainders in a coordinate system such that $\bF=\left(F,0\right)$
with $F>0$. To summarize, the parameters $b_{0}$ and $b_{1}$ are
zero because otherwise the torque $M$ is infinite and the parameter
$b_{2}$ of the third order is zero by our choice of coordinates.
Consequently, $a$ is the only free parameter of the asymptotic expansion
up to second order, and is related of the net force through \eqref{eq:F-a},
\begin{equation}
a=\left(\frac{F}{48}\right)^{1/3}\,.\label{eq:link-a-F}
\end{equation}

\section{Numerical simulations with standard and adaptive boundary conditions}

The aim of this section is to validate numerically the conjecture~\ref{conj:main}
concerning the existence of solutions satisfying \eqref{eq:ns-noforce}
and in particular the asymptotic expansion. We also provide a method
for solving this problem numerically in the spirit of \citet{Boenisch.etal-Adaptiveboundaryconditions2005,Boenisch.etal-Secondorderadaptive2008,Boeckle.Wittwer-Artificialboundaryconditions2013}
by using the asymptotes as an artificial boundary condition. To this
end we restrict the Navier-Stokes problem given by \eqref{eq:ns-noforce}
to an annulus $B(\bzero,R)\setminus B(\bzero,1)$ of radius $R$,
so that the Navier-Stokes system becomes
\begin{equation}
\begin{aligned}\Delta\bu-\bnabla p & =\bu\bcdot\bnabla\bu\,, & \bnabla\bcdot\bu & =0\,,\\
\left.\bu\right|_{\partial B(\bzero,1)} & =\bu^{*}\,, & \left.\bu\right|_{\Gamma} & =\bu_{\infty}\,,
\end{aligned}
\label{eq:ns-0-R}
\end{equation}
where $\Gamma=\partial B(\bzero,R)$ is the outer boundary, and $\bu_{\infty}$
is a priori the solution of the problem evaluated on $\Gamma$. Since
this solution is not known, a so called artificial boundary condition
has to be chosen.

\begin{figure}[H]
\includefigure{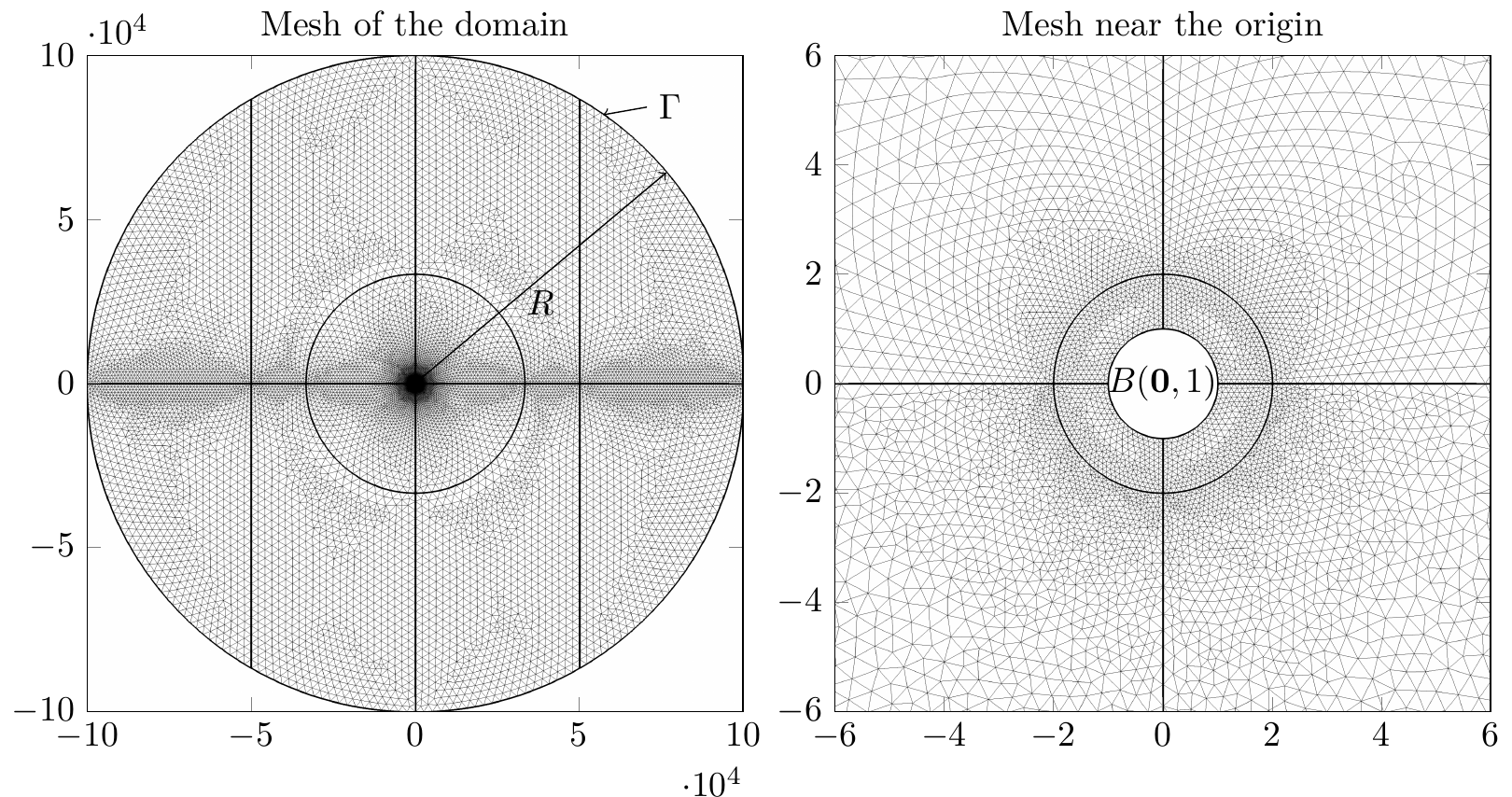}

\caption{\label{fig:mesh}Mesh of the domain used for the numerical simulations.
We refine the mesh along the horizontal line in order to capture more
accurately the decay in the wake region, and along the boundary of
$B(\bzero,1)$ to properly capture the boundary condition $\bu^{*}$.
The additional lines inside the domain are there to force the mesh
to have vertices on the lines where we will later on evaluate the
solution .}
\end{figure}

The simulations are done with COMSOL version 4.3, with the mesh presented
in figure~\ref{fig:mesh} and by using Lagrange P3 and P2 elements,
respectively, for the velocity and the pressure. In the numerical
simulations, we choose the boundary condition $\bu^{*}$ in the simplest
way that intuitively provides a net force: $\bu^{*}=\left(\lambda,0\right)$.
In the following we choose $\lambda=0.8$.

\subsection{No-slip boundary conditions}

In order to check the correctness of the asymptotic expansion, and
in view of the boundary condition at infinity in \eqref{eq:ns-0},
the simplest option for the artificial boundary condition is to choose
$\bu_{\infty}=\bzero$. Figure~\ref{fig:u-w-zero} shows the velocity
$u$ and the vorticity $\omega$ multiplied by the power of $r$ corresponding
to the decay of the solution predicted by the asymptotic expansion.
This way, we expect to see in the wake the functions of $z$ predicted
by the asymptotic expansion. The results confirm the decay of $u$
like $x^{-1/3}$ and of $\omega$ like $x^{-1}$. As expected, we
see that the velocity field, and especially also the vorticity, are
drastically modified within the wake near the artificial boundary
$\Gamma$.

\begin{figure}[H]
\includefigure{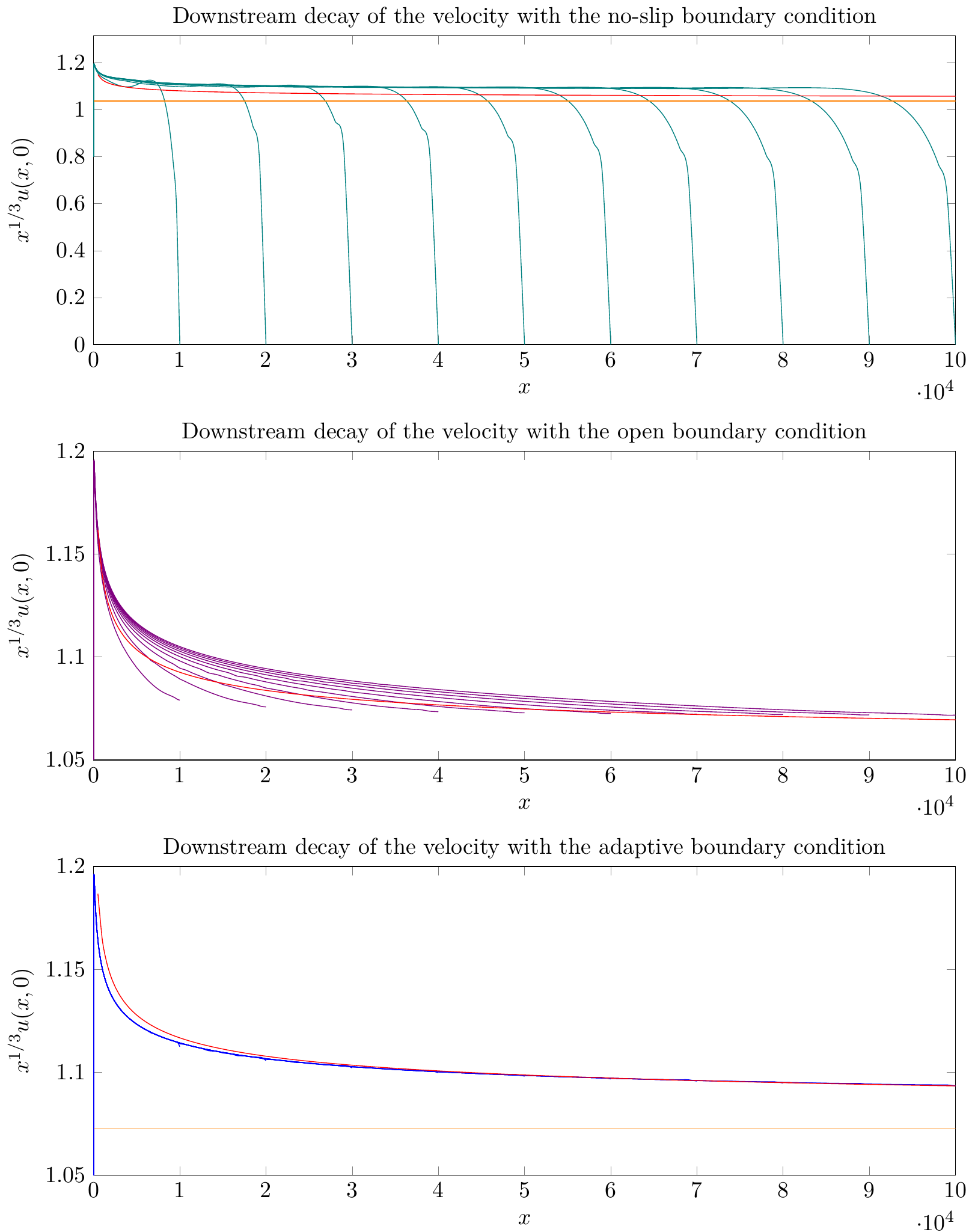}

\caption{\label{fig:u-down}Downstream velocity profile within the wake with
no-slip, open and adaptive boundary conditions on the outer boundary
$\Gamma$ for radii $R\in\left\{ 1,2,\dots,10\right\} \cdot10^{4}$.}
\end{figure}

In order to further validate the power law decay of $u$ and to check
the influence of the artificial boundary condition on the numerical
solution, we plot the velocity $u$ multiplied by $x^{1/3}$ downstream
and by $(-x)^{2/3}$ upstream on the line $y=0$, for different radii
$R$. Downstream (figure~\ref{fig:u-down}a), we see that the no-slip
boundary condition influences the velocity near the boundary $\Gamma$:
in a region of approximately constant size $10^{4}$ near the artificial
boundary, the solution depends on $R$, but, away from this region,
the simulations with different values of $R$ coincide, which is a
sign that the domain is big enough so that the artificial boundary
does not influence the solution in the centre of the domain. By measuring
the net force for the biggest domain, we deduce through \eqref{eq:link-a-F}
the value of $a$, and in the case $\lambda=0.8$, we find $a\approx0.416$.
Another method to determine $a$, is to minimize the integral
\begin{equation}
I(a)=\frac{4}{3\pi R^{2}}\int_{B(\mathbf{0},R)\setminus B(\mathbf{0},R/3)}r\left|\bu-\bu_{0}-\bu_{1}\right|\,,\label{eq:int-a}
\end{equation}
with respect to $a$. With this sort of fitting method, we find $a\approx0.418$.
In figure~\ref{fig:u-down}a, the orange line corresponds to the
main term at infinity,
\[
\lim_{x\to\infty}x^{1/3}u_{0}(x,0)=6a^{2}\,,
\]
and the red line to the asymptotic behaviour up to second order, 
\[
x^{1/3}\left(u_{0}(x,0)+u_{1}(x,0)\right)=6a^{2}+\frac{4a}{\sqrt{3}x^{1/3}}-\frac{\sqrt{3}}{\pi x^{2/3}}\,.
\]
The results confirm that the decay of the asymptotes and the relation
\eqref{eq:link-a-F} are correct.

Figure~\ref{fig:u-up}a concerns the upstream region, where the orange
line corresponds to the main term at infinity,
\[
\lim_{x\to-\infty}(-x)^{2/3}u_{0}(x,0)=\frac{4a}{\sqrt{3}}\,,
\]
and the red line to the asymptotic behaviour up to second order,
\[
(-x)^{2/3}\left(u_{0}(x,0)+u_{1}(x,0)\right)=\frac{4a}{\sqrt{3}}+\frac{\sqrt{3}}{\pi(-x)^{1/3}}\,.
\]
We see that the decay of $u$ is badly predicted with the no-slip
boundary condition, since the solutions for different values of $R$
coincide on a very small region only. The difficulty to predict the
upstream behaviour numerically is a known fact \citep[\S 4]{Boenisch.etal-Adaptiveboundaryconditions2005}.
Below, we provide a method which drastically improves the quality
of the results.

We conclude that the decay of the solution corresponds to the one
predicted by the asymptotic expansion. We now further analyse the
shape of the wake downstream and of the harmonic term upstream in
the largest domain, $R=10^{5}$, by restricting the solution to the
vertical lines $x=\pm R/2$, respectively. More precisely, in figure~\ref{fig:u-w-down-profile},
we plot $x^{1/3}u(x,x^{2/3}z)$ and $x\omega(x,x^{2/3}z)$ in terms
of $z=y/x^{2/3}$ at $x=R/2$. The red curve corresponds to the asymptotic
profile up to second order along the vertical line in consideration,
and the orange one to the leading profile only,
\begin{align*}
\lim_{x\to\infty}x^{1/3}u_{0}(x,x^{2/3}z) & =6a^{2}\sech^{2}(az)\,,\\
\lim_{x\to\infty}x^{1/3}\omega_{0}(x,x^{2/3}z) & =-12a^{3}\sinh(az)\sech^{3}(az)\,.
\end{align*}
The agreement of the simulations for both velocity and vorticity with
the asymptotic expansion is very good. To check the profile upstream,
we plot in figure~\ref{fig:u-w-up-profile} $(-x)^{2/3}u(x,xz)$
and $x^{2}\omega(x,xz)$ in terms of $z=-y/x$ with $x=-R/2$. The
red curves are again the asymptotic profile up to second order and
the orange one, the leading term,
\begin{eqnarray*}
\lim_{x\to-\infty}(-x)^{2/3}u_{0}(x,xz) & = & \frac{4a}{\sqrt{3}}\frac{\cos(\arctan(z)/3)+z\sin(\arctan(z)/3)}{\left(1+z^{2}\right)^{5/6}}\,,\\
\lim_{x\to-\infty}x^{2}\omega_{0}(x,xz) & = & 0\,.
\end{eqnarray*}
As already observed for the decay of $u$, the profiles are less good
upstream than downstream. Again, we will provide below a method which
improves especially the upstream profile.

\subsection{Open boundary conditions}

\begin{figure}[H]
\includefigure{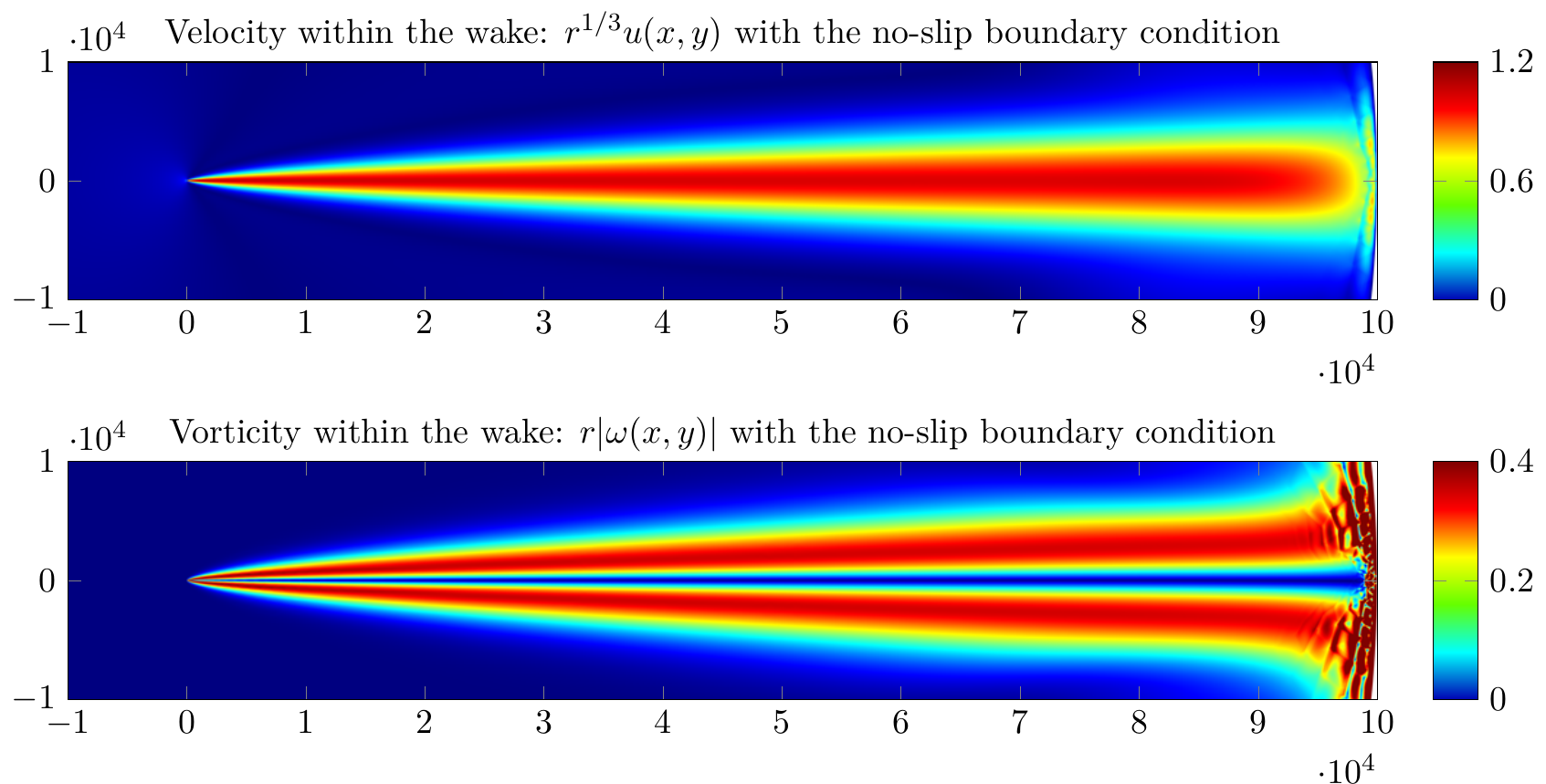}

\caption{\label{fig:u-w-zero}Velocity and vorticity profiles with the no-slip
boundary condition in an annulus of radius $R=10^{5}$. The wake for
$u$ has the form $u\approx x^{-1/3}\varphi_{0}^{\prime}(z)$ and
the one of the vorticity $\omega\approx-x^{-1}\varphi_{0}^{\prime\prime}(z)$.}
\end{figure}

\begin{figure}[H]
\includefigure{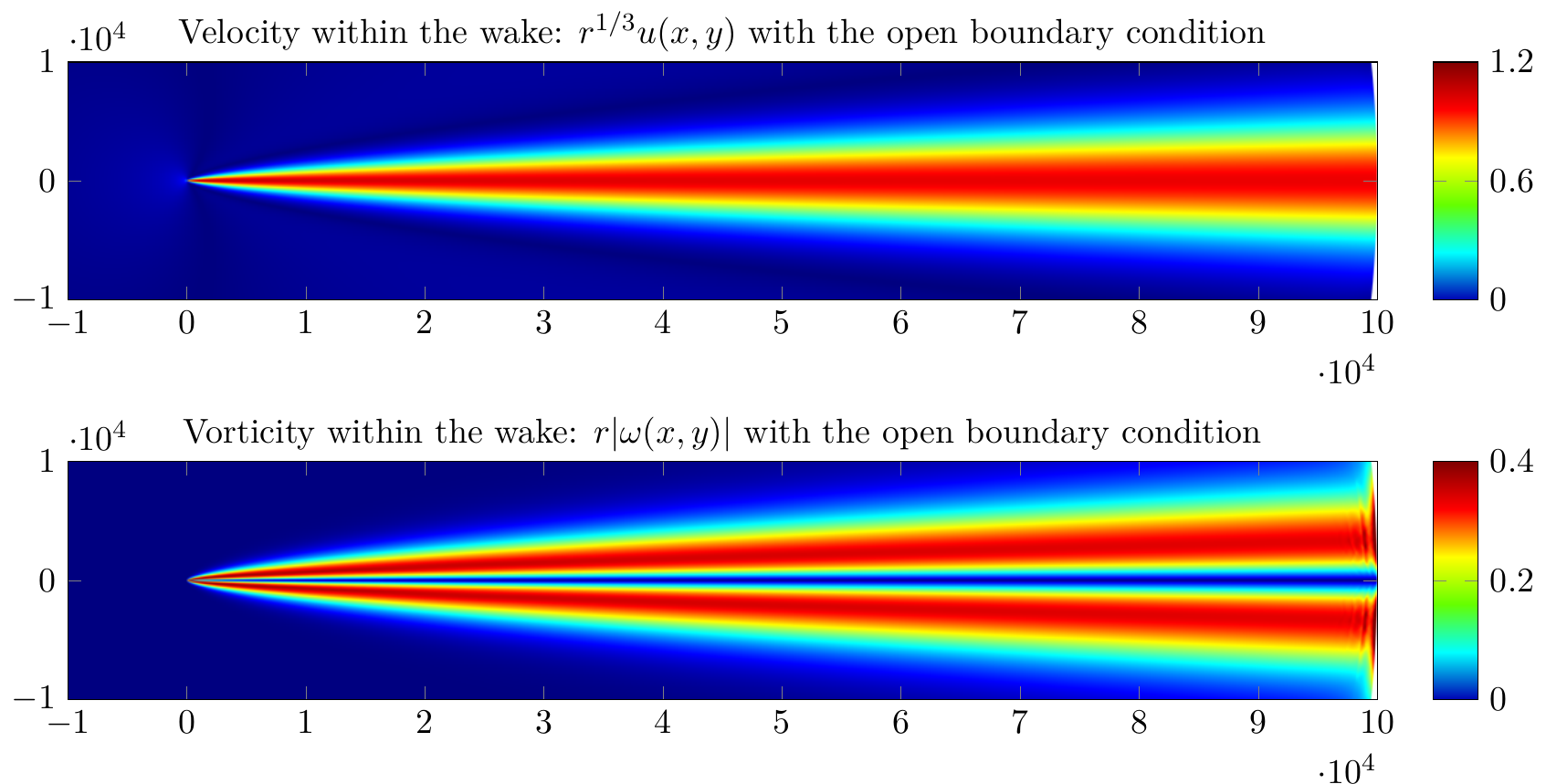}

\caption{\label{fig:u-w-slip}Velocity and vorticity profiles with the open
boundary condition in an annulus of radius $R=10^{5}$. The wake is
less influenced near the artificial boundary by the open boundary
condition than by the no-slip boundary condition.}
\end{figure}

One standard way to improve the simulations near the artificial boundary
$\Gamma$, is to choose an artificial boundary condition that does
not fix the value of $\bu$ on the boundary, but rather a quantity
related to the derivatives. Here we choose the so called open boundary
condition,
\[
\left[\bnabla\bu+\left(\bnabla\bu\right)^{T}-p\right]\bn=\bzero\quad\text{on}\;\Gamma\,.
\]
Figure~\ref{fig:u-w-slip} is the analogue of figure~\ref{fig:u-w-zero}
but with the open artificial boundary condition. From a qualitative
point of view, the wake for $u$ is almost not influenced by the open
artificial boundary condition. For the vorticity $\omega$, we see
some small deviations in the wake near the artificial boundary. This
is expected, since the open boundary condition fixes quantities related
to the derivative of $\bu$ and the vorticity $\omega$ is defined
trough derivatives of $\bu$. Figure~\ref{fig:u-down}b, shows the
decay of $u$ along the $x$-axis, and we see that the open boundary
condition is better suited to predict the downstream behaviour. Figure~\ref{fig:u-up}b
exhibits that the upstream behaviour is as badly predicted as with
the no-slip boundary condition. Finally, the velocity and the vorticity
profiles downstream (figure~\ref{fig:u-w-down-profile}) are rather
not influenced by the choice of the boundary condition. Again the
upstream profiles (figure~\ref{fig:u-w-up-profile}) are not in good
agreement with the asymptotic expansion. We conclude that the open
boundary condition improves the simulations downstream, but not upstream.
The aim of the next subsection is to show that, when using the asymptotic
expansion to define artificial boundary conditions, the results of
the simulations are vastly improved in the upstream region.

\subsection{Adaptive boundary conditions}

\begin{figure}[H]
\includefigure{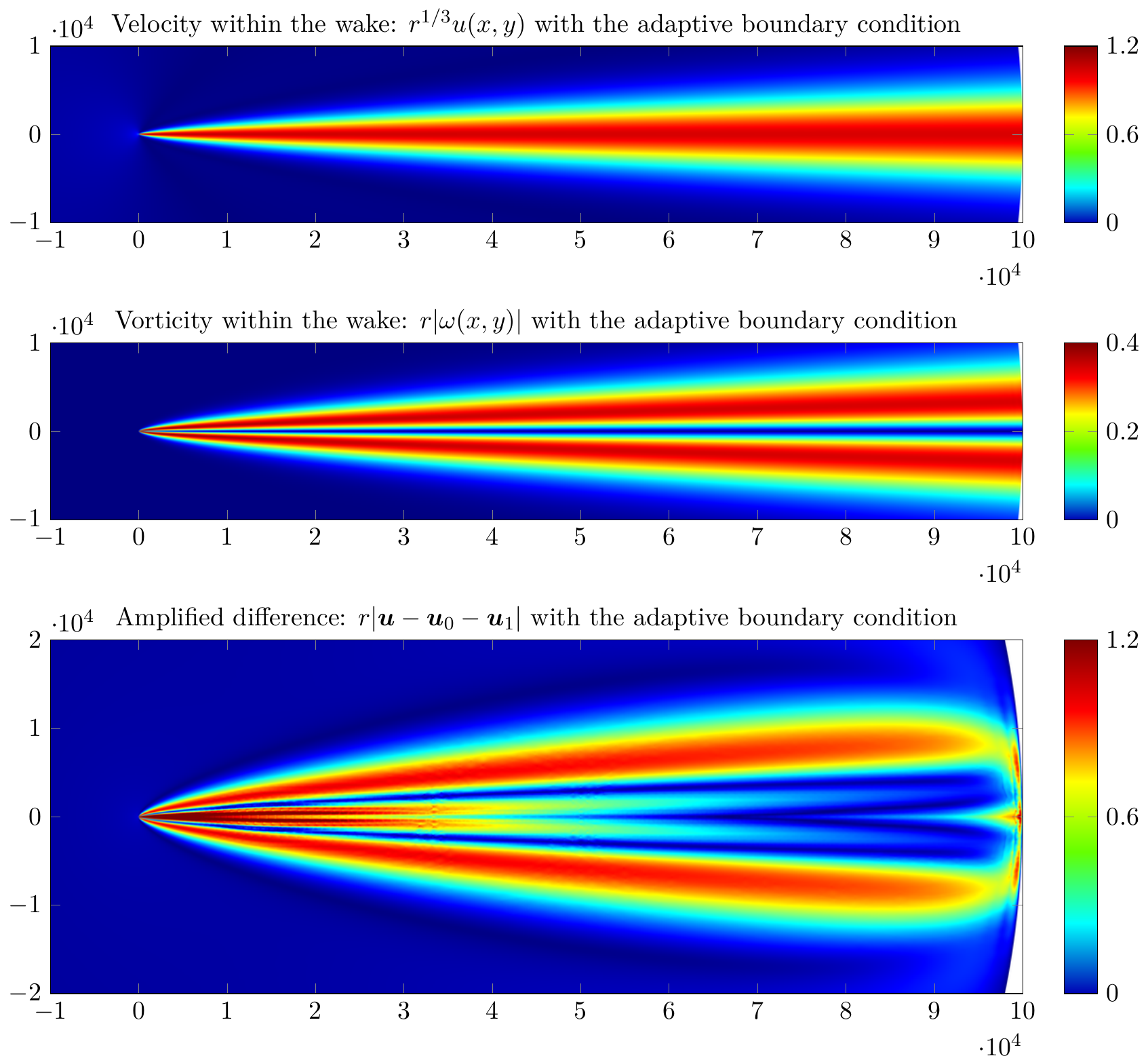}

\caption{\label{fig:u-w-asy}Velocity and vorticity profiles with the adaptive
boundary condition. Difference between the numerical velocity and
the asymptotic expansion: the dominant remaining term is of the form
$\left|\bu-\bu_{0}-\bu_{1}\right|\approx x^{-1}\varphi_{2}^{\prime\prime}(z)$
as expected from the computation of the third order of the asymptotic
expansion.}
\end{figure}

\begin{figure}[H]
\includefigure{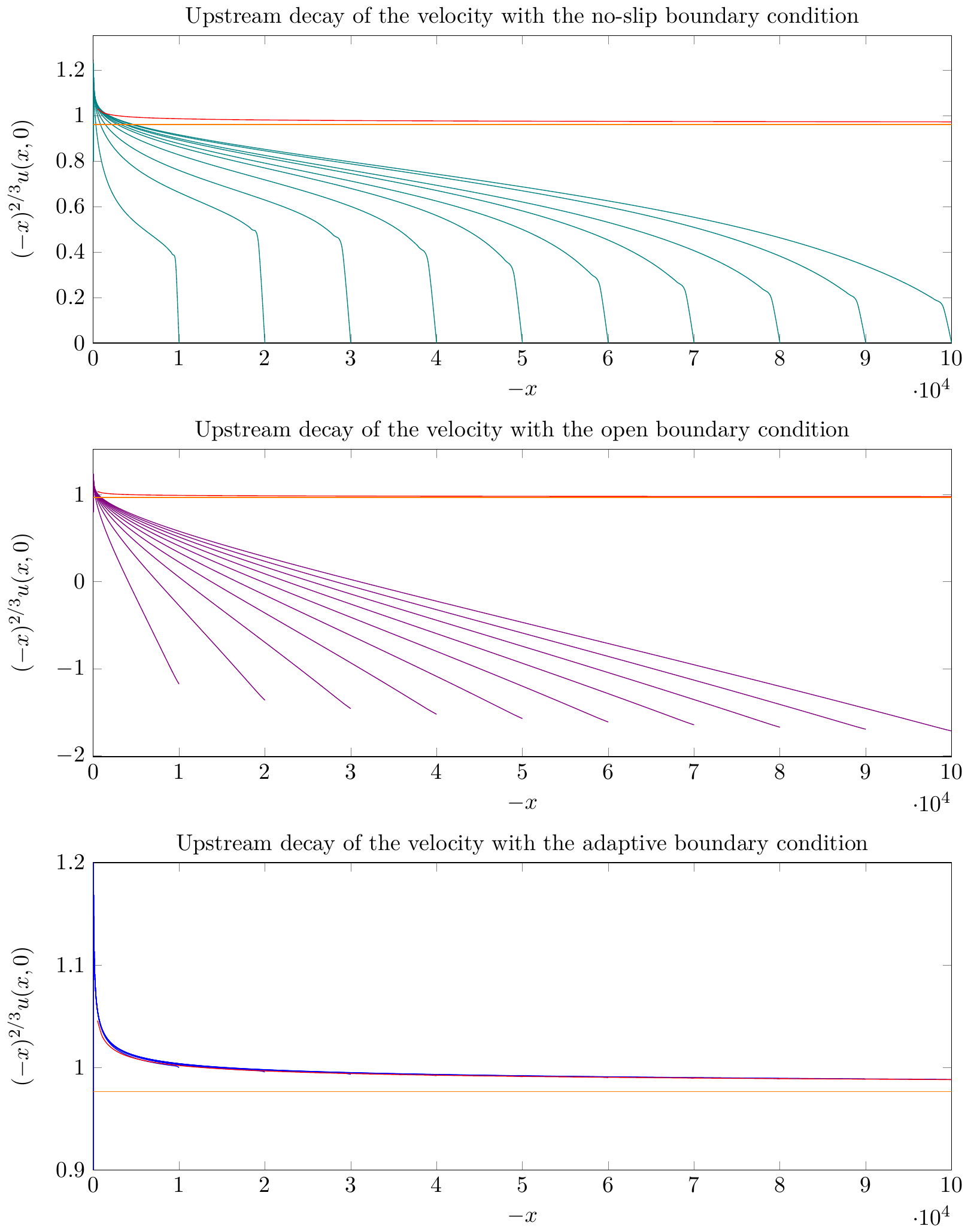}

\caption{\label{fig:u-up}Upstream velocity profile with no-slip, open and
adaptive boundary conditions on the outer boundary $\Gamma$ for radii
$R\in\left\{ 1,2,\dots,10\right\} \cdot10^{4}$.}
\end{figure}

\begin{figure}[H]
\includefigure{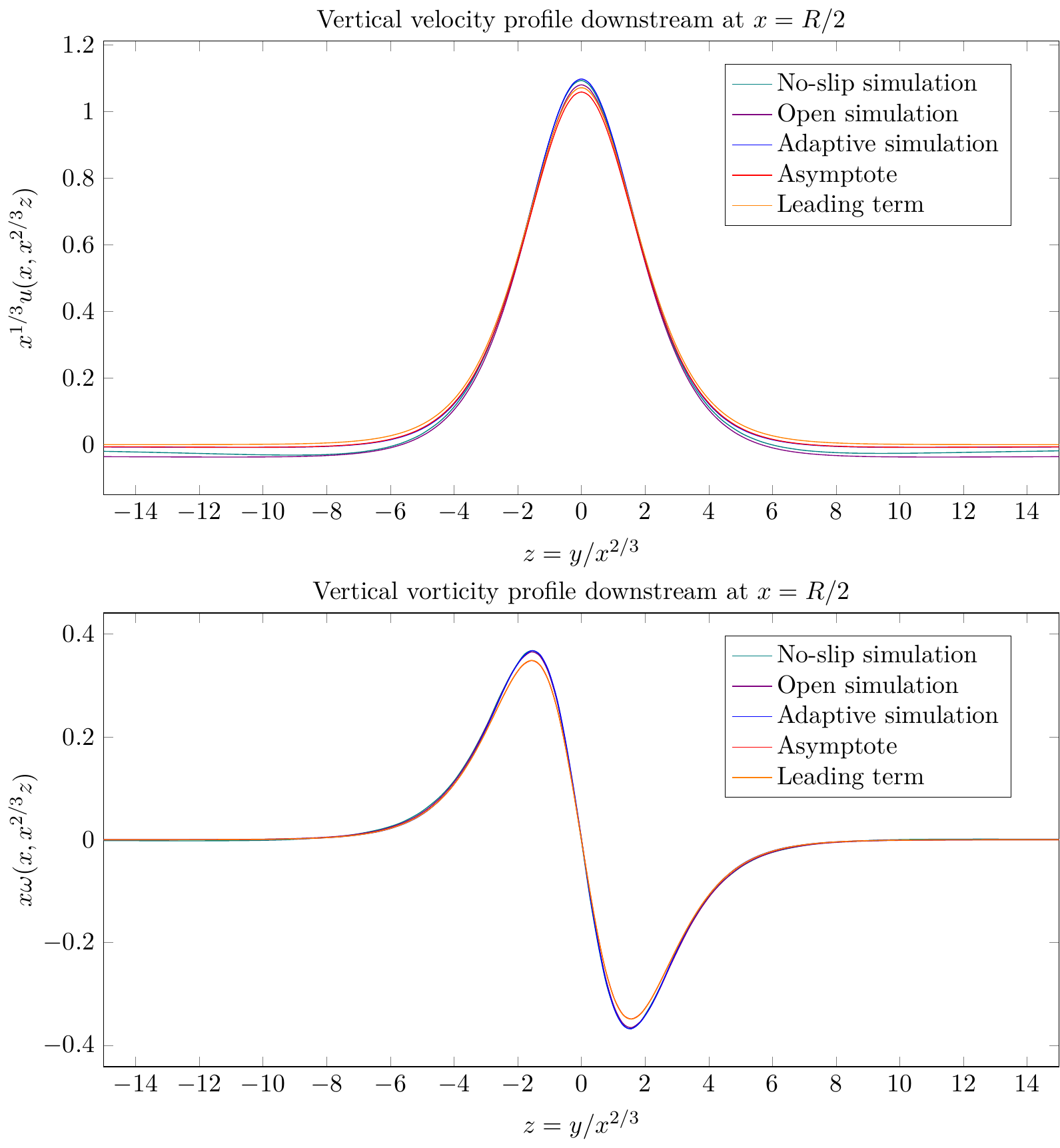}

\caption{\label{fig:u-w-down-profile}Velocity and vorticity profiles along
the vertical line at $x=R/2$ with $R=10^{5}$ in terms of $z=y/x^{2/3}$.}
\end{figure}

\begin{figure}[H]
\includefigure{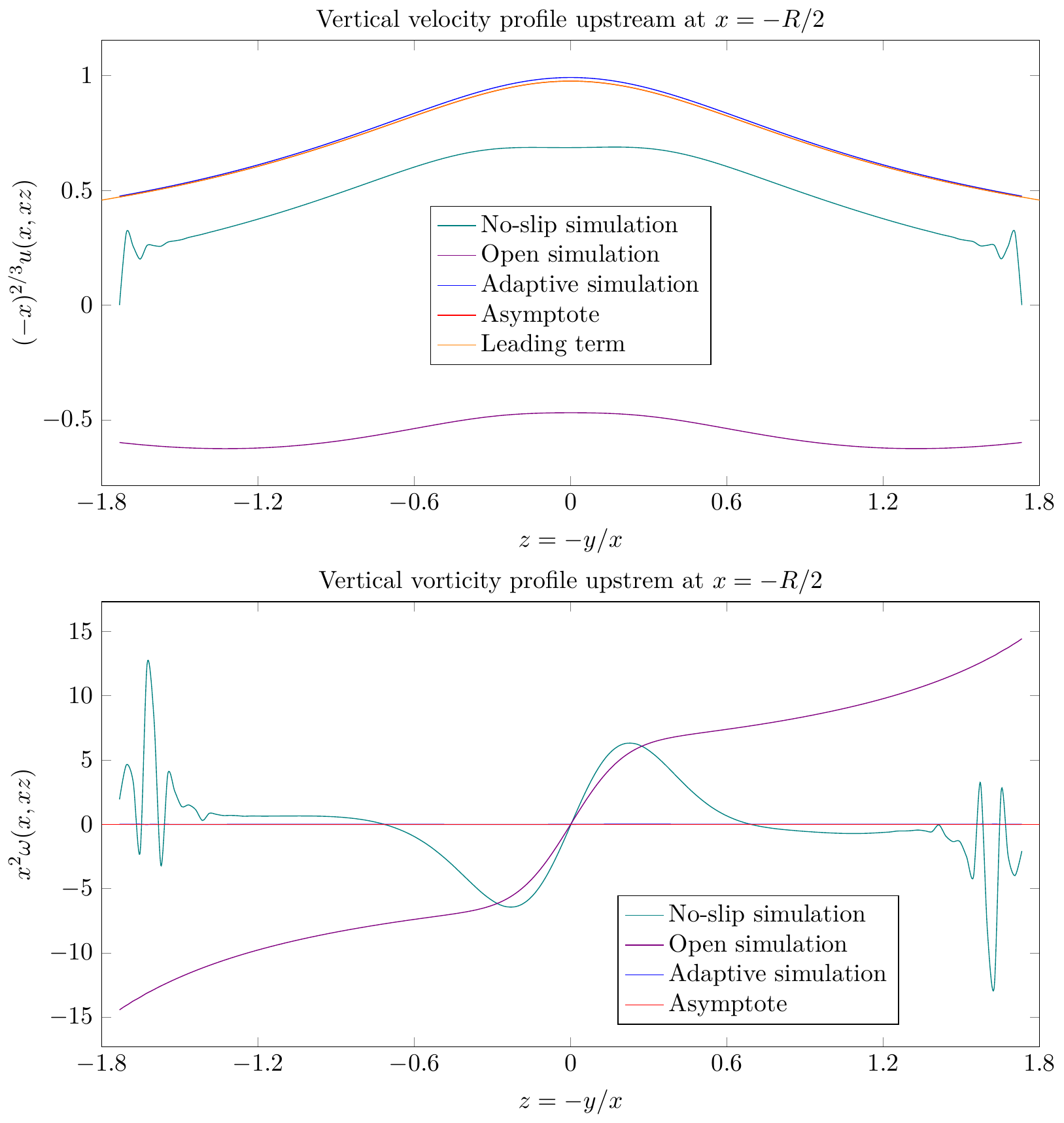}

\caption{\label{fig:u-w-up-profile}Velocity and vorticity profiles along the
vertical line at $x=-R/2$ with $R=10^{5}$ in terms of $z=-y/x$.
As expected, the vorticity in almost zero in the upstream region.}
\end{figure}

Since the asymptotic expansion provides information on the behaviour
of the solution at large values of $r$, it is natural to evaluate
the asymptotic expansion on the artificial boundary,
\[
\bu_{\infty}=\left.\bu_{0}\right|_{\Gamma}+\left.\bu_{1}\right|_{\Gamma}\,,
\]
instead of taking the value at infinity as the artificial boundary
condition. The asymptotic expansion depends on the free parameter
$a$, which is related to the net force acting on the body. Since
the force is not known before the solution is computed, we determine
$a$ by minimizing the integral $I(a)$ defined by \eqref{eq:int-a}
with respect to $a$. This is done numerically by using the SNOPT
algorithm of COMSOL, which is a gradient method. We note that this
step is not needed for the case of problem \eqref{eq:ns-force} since
the net force \eqref{eq:FM-domain} is known before the solution is
computed. In the case $\lambda=0.8$, we find $a\approx0.423$. As
shown in figure~\ref{fig:u-w-asy}a\&b, the qualitative behaviour
of the wake near the artificial boundary seems to be insensitive to
the artificial boundary, both for the velocity and the vorticity.
In figure~\ref{fig:u-w-asy}c, we plot the difference between the
velocity and the asymptotic expansion up to second order. This difference
appears to decay like $r^{-1}$ as predicted by conjecture~\ref{conj:main},
and corresponds to the third order term $\psi_{2}=x^{-1/3}\varphi_{2}(z)$
in the asymptotic expansion \eqref{eq:expansion}. With the adaptive
boundary condition, the decay of $u$ downstream (figure~\ref{fig:u-down}c)
and upstream (figure~\ref{fig:u-up}c) is not modified by the size
of the domain. We conclude that the solution can be accurately computed
with this type of boundary conditions on much smaller domains than
with no-slip or open boundary conditions. As can be seen, the vertical
profiles downstream (figure~\ref{fig:u-w-down-profile}) and upstream
(figure~\ref{fig:u-w-up-profile}) are so close to the second order
asymptotes that the two lines essentially become indistinguishable.
We take these results as a further confirmation of the correctness
of the asymptotic expansion.

One of the big advantage of adaptive boundary conditions over standard
boundary conditions, is that the upstream behaviour is also captured
accurately. This allows to take domains much smaller without modifying
the quality of the simulations, and to reduce the computational time,
in spite of the disadvantage that one needs to perform multiple simulations
with different values of $a$ in order to find the minimum of $I(a)$.
The reason why, this techinque is not that time consuming is that
the solver profits from a good initial guess in the non-linear iterations
by taking the solution already computed for a nearby value of $a$.

Finally, we also analyse the solutions for smaller values of $\lambda$.
As shown in movie~\href{http://yaros.unige.ch/GuillodWittwer-2013/movie1/}{1},
by taking $\lambda$ smaller, the parameter $a$ and the net force
get also smaller, so that the amplitude of the wake decreases, and
the wake becomes wider.

\section{Numerical simulations with rotating boundary conditions}

\begin{figure}[H]
\includefigure{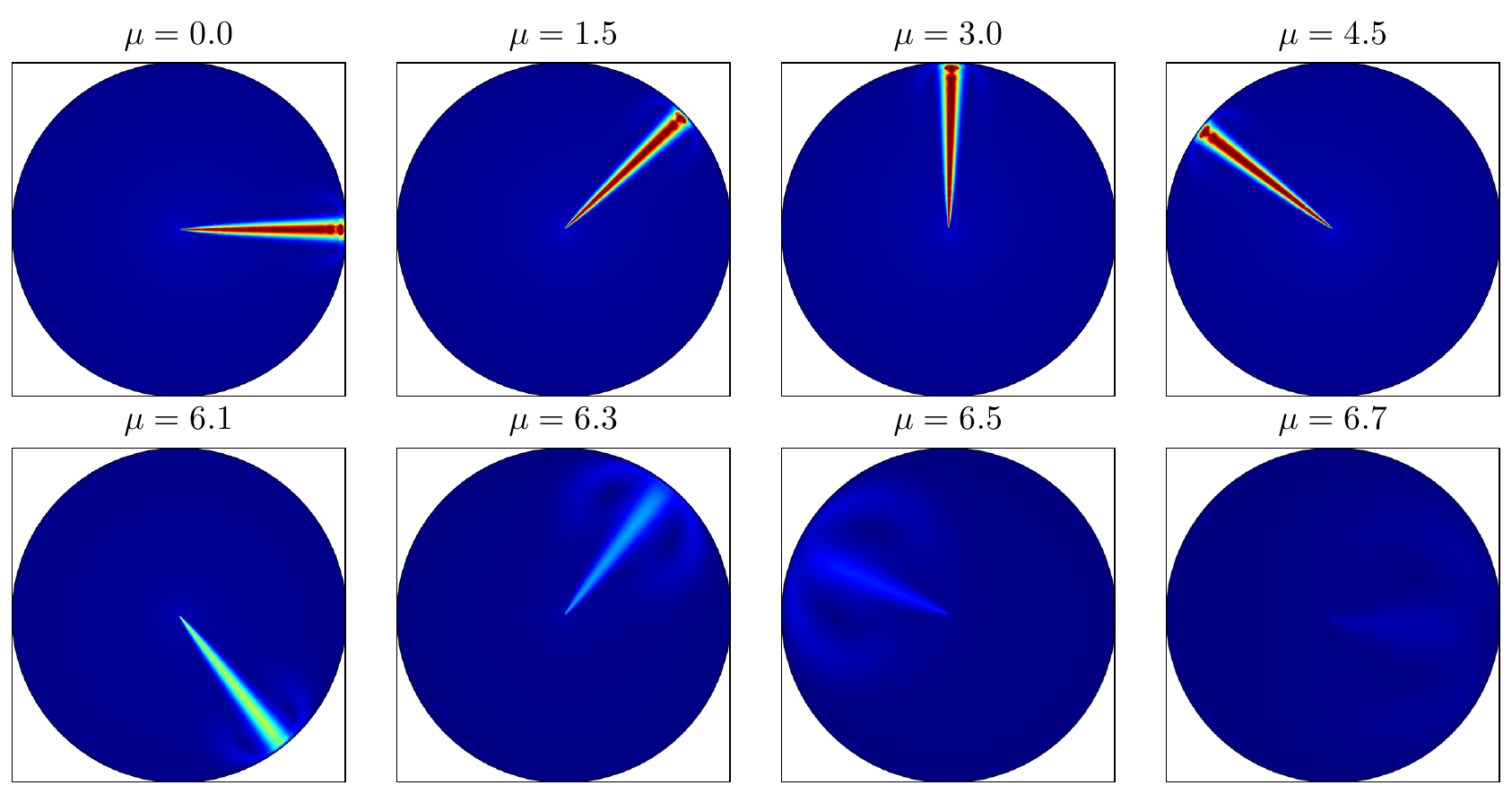}

\caption{\label{fig:mu-profile}Magnitude of the velocity field $|\bx|^{1/3}|\bu|$
for different values of $\mu$ with $R=10^{5}$. As $\mu$ grows the
wake starts to rotate and its amplitude decays, and for $\mu>\sqrt{48}$
the wake has totally disappeared.}
\end{figure}

\begin{figure}[H]
\includefigure{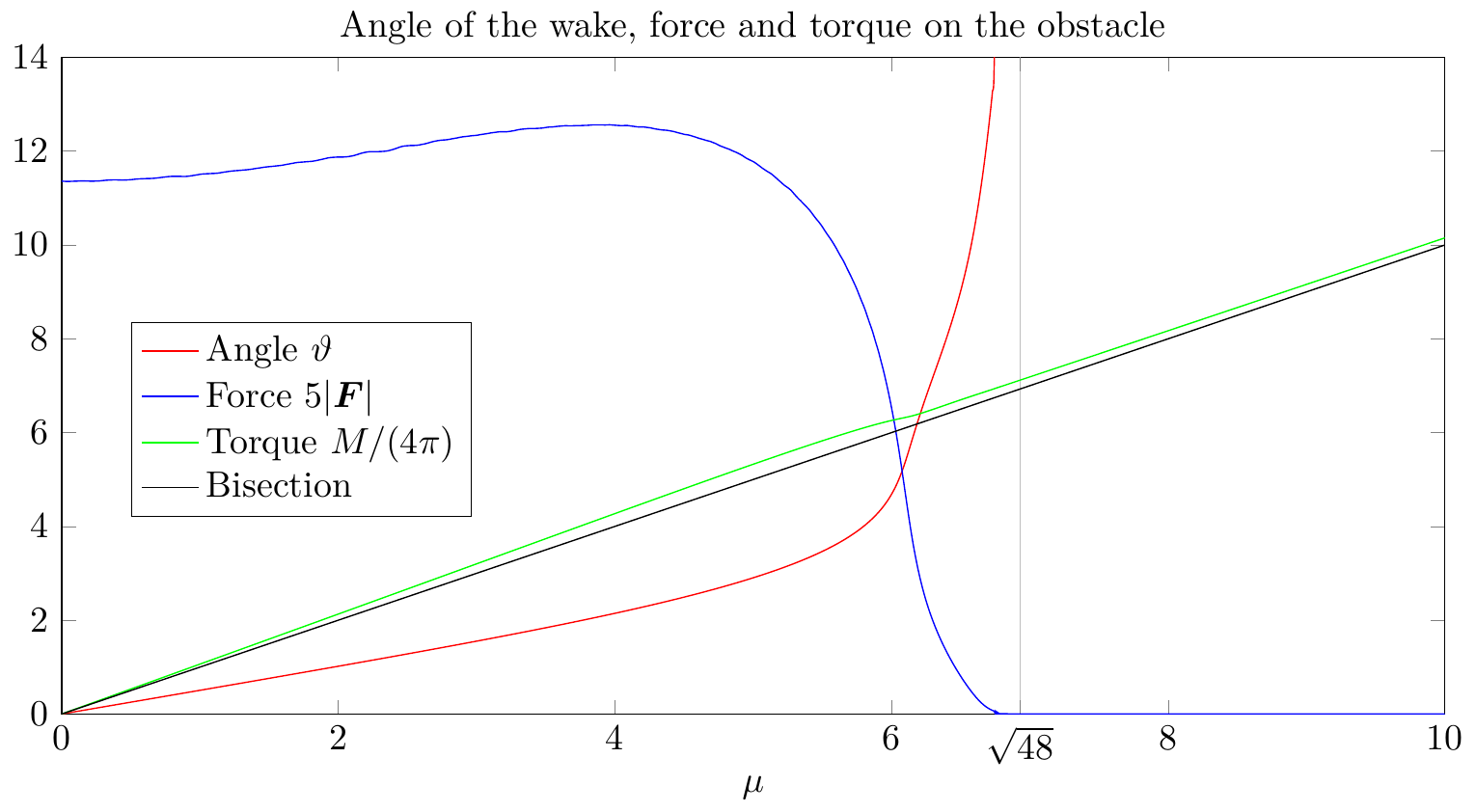}

\caption{\label{fig:mu-force}Magnitude of the force and torque acting on the
boundary $\partial B(\bzero,1)$ and angle $\vartheta$ of the orientation
of the wake as a function of $\mu$ for $R=10^{5}$. For $\mu>\sqrt{48}$
the wake is no more present, so no angle is measured.}
\end{figure}

Recently, \citet{Hillairet-mu2013} proved the existence of solutions
to \eqref{eq:ns-noforce} for a ball of boundary conditions which
contains in particular the function
\[
\bu^{*}=\mu\left(-y,x\right)+\lambda\left(1,0\right)\,,
\]
for the case where $\mu>\sqrt{48}$ and $\lambda$ is sufficiently
small. The authors prove the existence of a solution decaying uniformly
like $r^{-1}$, and more precisely that the asymptotic behaviour is
given by the stream function $\psi=\tilde{\mu}\log r$ where $\tilde{\mu}$
is close to $\mu$. The aim of this section is to analyse the transition
from $\mu=0$, which corresponds to the case previously considered,
to $\mu>\sqrt{48}$ where there is no net force. For the numerical
simulations, we take $\lambda=0.8$ as before and use the parametric
solver of COMSOL to vary the parameter $\mu$. Figure~\ref{fig:mu-profile}
and movie~\href{http://yaros.unige.ch/GuillodWittwer-2013/movie2/}{2}
show the wake for the velocity for different values of $\mu$ with
no-slip boundary condition, $\bu_{\infty}=\bzero$. We find that the
more $\mu$ is increasing the more the wake is rotated, but the shape
remains otherwise unchanged. Just before $\mu=\sqrt{48}$, the wake
amplitude is starting to decrease and the orientation changes more
rapidly, so that in a very few numerical steps of $\mu$ the wake
has totally disappeared. Figure~\ref{fig:mu-force} shows, as a function
of $\mu$, the angle $\vartheta$ between the wake and the positive
real axis as well as the force acting on the body. The angle varies
linearly when $\mu$ is small and apparently diverges when $\mu$
approaches $\sqrt{48}$. For $\mu>\sqrt{48}$, the wake is no more
present, and the force is zero as expected. We also compute the torque
$M$ on the body measured from its centre, which appears to be linearly
increasing with $\mu$. The slope is approximately given by $4\pi$
which is the value analytically obtained in the case $\lambda=0$.





\bibliographystyle{jfm}
\phantomsection\addcontentsline{toc}{section}{\refname}\bibliography{paper}

\begin{thebibliography}{30}
\expandafter\ifx\csname natexlab\endcsname\relax\def\natexlab#1{#1}\fi

\bibitem[Babenko(1973)]{Babenko-stationarysolutionsof1973}
{\sc Babenko, K.~I.} 1973 On stationary solutions of the problem of flow past a
  body by a viscous incompressible fluid. {\em Math. USSR-Sb.\/} {\bf 20},
  1--25.

\bibitem[Bichsel \& Wittwer(2007)]{Bichsel.Wittwer-Stationaryflowpast2007}
{\sc Bichsel, D. \& Wittwer, P.} 2007 Stationary flow past a semi-infinite flat
  plate: analytical and numerical evidence for a symmetry-breaking solution.
  {\em Journal of Statistical Physics\/} {\bf 127}~(1), 133--170.

\bibitem[Boeckle \&
  Wittwer(2013)]{Boeckle.Wittwer-Artificialboundaryconditions2013}
{\sc Boeckle, C. \& Wittwer, P.} 2013 Artificial boundary conditions for
  stationary {N}avier-{S}tokes flows past bodies in the half-plane. {\em
  Computers \& Fluids\/} {\bf 82}, 95--109.

\bibitem[B{\"o}nisch {\em et~al.\/}(2005)B{\"o}nisch, Heuveline \&
  Wittwer]{Boenisch.etal-Adaptiveboundaryconditions2005}
{\sc B{\"o}nisch, S., Heuveline, V. \& Wittwer, P.} 2005 Adaptive boundary
  conditions for exterior flow problems. {\em Journal of Mathematical Fluid
  Mechanics\/} {\bf 7}~(1), 85--107.

\bibitem[B{\"o}nisch {\em et~al.\/}(2008)B{\"o}nisch, Heuveline \&
  Wittwer]{Boenisch.etal-Secondorderadaptive2008}
{\sc B{\"o}nisch, S., Heuveline, V. \& Wittwer, P.} 2008 Second order adaptive
  boundary conditions for exterior flow problems: Non-symmetric stationary
  flows in two dimensions. {\em Journal of Mathematical Fluid Mechanics\/} {\bf
  10}~(1), 45--70.

\bibitem[Deuring \& Galdi(2000)]{Deuring.Galdi-AsymptoticBehaviorof2000}
{\sc Deuring, P. \& Galdi, G.~P.} 2000 On the asymptotic behavior of physically
  reasonable solutions to the stationary {N}avier-{S}tokes system in
  three-dimensional exterior domains with zero velocity at infinity. {\em
  Journal of Mathematical Fluid Mechanics\/} {\bf 2}, 353--364.

\bibitem[van Dyke(1975)]{Dyke-Perturbationmethodsin1975}
{\sc van Dyke, M.} 1975 {\em Perturbation methods in fluid mechanics\/},
  annotated edn. Stanford, California: The Parabolic Press.

\bibitem[Farwig \& Sohr(1998)]{Farwig.Sohr-WeightedestimatesOseen1998}
{\sc Farwig, R. \& Sohr, H.} 1998 Weighted estimates for the {O}seen equations
  and the {N}avier-{S}tokes equations in exterior domains. Heywood, J. G. (ed.)
  et al., Theory of the Navier-Stokes equations. Proceedings of the third
  international conference on the Navier-Stokes equations: theory and numerical
  methods, Oberwolfach, Germany, June 5--11, 1994. Singapore: World Scientific.
  Ser. Adv. Math. Appl. Sci. 47, 11--30.

\bibitem[Finn(1959)]{Finn-Estimatesatinfinity1959}
{\sc Finn, R.} 1959 Estimates at infinity for stationary solutions of the
  {N}avier-{S}tokes equations. {\em Bull. Math. Soc. Sci. Math. Phys. R. P.
  Roumaine (N.S.)\/} {\bf 3 (51)}, 387--418.

\bibitem[Finn \& Smith(1967)]{Finn-stationarysolutionsNavier1967}
{\sc Finn, R. \& Smith, D.~R.} 1967 On the stationary solutions of the
  {N}avier-{S}tokes equations in two dimensions. {\em Archive for Rational
  Mechanics and Analysis\/} {\bf 25}, 26--39.

\bibitem[Galdi(1992)]{Galdi-AsymptoticStructureDsol-1992}
{\sc Galdi, G.~P.} 1992 On the asymptotic structure of $d$-solutions to steady
  {N}avier-{S}tokes equations in exterior domains. In {\em Mathematical
  problems relating to the Navier-Stokes equation\/} (ed. G.~P. Galdi), {\em
  Series on advances in mathematics for applied sciences\/}, vol.~11, pp.
  81--105. World Scientific.

\bibitem[Galdi(1993)]{Galdi-asymptoticPropertiesLerays1993}
{\sc Galdi, G.~P.} 1993 On the asymptotic properties of {L}eray's solutions to
  the exterior steady three-dimensional {N}avier-{S}tokes equations with zero
  velocity at infinity. In {\em Degenerate Diffusions\/} (ed. Wei-Ming Ni,
  L.~A. Peletier \& J.~L. Vazquez), {\em The IMA Volumes in Mathematics and its
  Applications\/}, vol.~47, pp. 95--103. Springer.

\bibitem[Galdi(2004)]{Galdi-StationaryNavier-Stokesproblem2004}
{\sc Galdi, G.~P.} 2004 Stationary {N}avier-{S}tokes problem in a
  two-dimensional exterior domain. In {\em Stationary partial differential
  equations. Vol. I\/}, pp. 71--155. Amsterdam: North-Holland.

\bibitem[Galdi(2011)]{Galdi-IntroductiontoMathematical2011}
{\sc Galdi, G.~P.} 2011 {\em An Introduction to the Mathematical Theory of the
  {N}avier-{S}tokes Equations. Steady-State Problems\/}, second edition edn.
  New York: Springer Verlag.

\bibitem[Galdi \& Sohr(1995)]{Galdi.Sohr-asymptoticstructureof1995}
{\sc Galdi, G.~P. \& Sohr, H.} 1995 On the asymptotic structure of plane steady
  flow of a viscous fluid in exterior domains. {\em Archive for Rational
  Mechanics and Analysis\/} {\bf 131}, 101--119.

\bibitem[Goldstein(1957)]{Goldstein-LecturesFluidMechanics1957}
{\sc Goldstein, S.} 1957 {\em Lectures on Fluid Mechanics\/}. Interscience
  Publishers, Ltd., London.

\bibitem[Hamel(1917)]{Hamel-SpiralfoermigeBewegungen1917}
{\sc Hamel, G.} 1917 {S}piralf\"{o}rmige {B}ewegungen z\"{a}her
  {F}l\"{u}ssigkeiten. {\em Jahresbericht der Deutschen
  Mathematiker-Vereinigung\/} {\bf 25}, 34--60.

\bibitem[Hillairet \&
  Wittwer(2012)]{Hillairet.Wittwer-Asymptoticdescriptionof2011}
{\sc Hillairet, M. \& Wittwer, P.} 2012 Asymptotic description of solutions of
  the exterior {N}avier-{S}tokes problem in a half space. {\em Archive for
  Rational Mechanics and Analysis\/} {\bf 205}, 553--584.

\bibitem[Hillairet \& Wittwer(2013)]{Hillairet-mu2013}
{\sc Hillairet, M. \& Wittwer, P.} 2013 On the existence of solutions to the
  planar exterior {N}avier-{S}tokes system.

\bibitem[Korolev \&
  \v{S}ver\'ak(2011)]{Korolev.Sverak-largedistanceasymptotics2011}
{\sc Korolev, A. \& \v{S}ver\'ak, V.} 2011 On the large-distance asymptotics of
  steady state solutions of the {N}avier-{St}okes equations in {3D} exterior
  domains. {\em Annales de l'Institut Henri Poincar{\'{e}} - Analyse non
  lin{\'{e}}aire\/} {\bf 28}~(2), 303--313.

\bibitem[Landau(1944)]{Landau-newexactsolution1944}
{\sc Landau, L.~D.} 1944 A new exact solution of the {N}avier-{S}tokes
  equations. {\em Doklady Akademii Nauk SSSR\/} {\bf 43}, 286--288.

\bibitem[Nazarov \& Pileckas(2000)]{Nazarov-steady2000}
{\sc Nazarov, S.~A. \& Pileckas, K.} 2000 On steady {S}tokes and
  {N}avier-{S}tokes problems with zero velocity at infinity in a
  three-dimensional exterior domain. {\em Journal of Mathematics of Kyoto
  University\/} {\bf 40}, 475--492.

\bibitem[Nazarov \& Pileckas(1999)]{Nazarov.Pileckas-AsymptoticofSolutions1999}
{\sc Nazarov, S.~A. \& Pileckas, K.~I.} 1999 Asymptotic of solutions of the
  {N}avier-{S}tokes equations in the exterior of a bounded body. {\em Doklady
  Mathematics\/} {\bf 60}~(1), 133--135.

\bibitem[Ockendon \& Ockendon(1995)]{Ockendon.Ockendon-ViscousFlow1995}
{\sc Ockendon, H. \& Ockendon, J.~R.} 1995 {\em Viscous Flow\/}. Cambridge
  University Press.

\bibitem[Pileckas \& Russo(2012)]{Pileckas-existencevanishing2012}
{\sc Pileckas, K. \& Russo, R.} 2012 On the existence of vanishing at infinity
  symmetric solutions to the plane stationary exterior {N}avier-{S}tokes
  problem. {\em Mathematische Annalen\/} {\bf 352}~(3), 643--658.

\bibitem[Russo(2008)]{Russo-Steady-StateNavier-StokesEquations2008}
{\sc Russo, A.} 2008 On the steady-state {N}avier-{St}okes equations in two
  dimensional domains. PhD thesis, Universit\`{a} di Napoli "Federico {II}".

\bibitem[Sazonov(1999)]{Sazonov-AsymptoticBehaviorof1999}
{\sc Sazonov, L.~I.} 1999 Asymptotic behavior of the solution to the
  two-dimensional stationary problem of flow past a body far from it. {\em
  Mathematical Notes\/} {\bf 65}~(2), 202--207.

\bibitem[\v{S}ver\'ak(2011)]{Sverak-LandausSolutionsNavier2011}
{\sc \v{S}ver\'ak, V.} 2011 On {L}andau's solutions of the {N}avier-{S}tokes
  equations. {\em Journal of Mathematical Sciences\/} {\bf 179}~(1), 208--228,
  translated from \textit{Problems in Mathematical Analysis} \textbf{61},
  October 2011, pp. 175-190.

\bibitem[Yamazaki(2009)]{Yamazaki-stationaryNavier-Stokesequation2009}
{\sc Yamazaki, M.} 2009 The stationary {N}avier-{S}tokes equation on the whole
  plane with external force with antisymmetry. {\em Annali dell'Universita di
  Ferrara\/} {\bf 55}, 407--423.

\bibitem[Yudovich(2003)]{Yudovich-ElevenGreatProblems2003}
{\sc Yudovich, V.~I.} 2003 Eleven great problems of mathematical hydrodynamics.
  {\em Moscow Mathematical Journal\/} {\bf 3}~(2), 711--737.

\end{thebibliography}

\end{document}